\documentclass{JHEP}
\input epsf.tex

\usepackage[caption=false]{subfig}

\epsfclipon

\usepackage{epsfig}
\usepackage{epstopdf}
\usepackage{epsf}
\input{epsf.sty}
\usepackage{graphicx,amsmath,amssymb}
\usepackage{epsfig,multicol}

\usepackage{bbm,bm,amsmath,amssymb}
\def\bg{\begin{eqnarray}}
\def\nd{\end{eqnarray}}
\def\be{\begin{equation}}
\def\ee{\end{equation}}
\def\ba{\begin{eqnarray}}
\def\ea{\end{eqnarray}}

\newcommand{\roughly}[1]{\mathrel{\raise.3ex\hbox{$#1$\kern-0.85em
\lower1ex\hbox{$\sim$}}}}

\def\2pi{\left(2\pi\right)}

\def\beq{\begin{equation}}
\def\eeq{\end{equation}}
\def\beqa{\begin{eqnarray}}
\def\eeqa{\end{eqnarray}}
\def\bea{\begin{eqnarray}}
\def\eea{\end{eqnarray}}

\def\D3{\overline{\mbox{D3}}}

\title{Non-extremal geometries and holographic phase transitions}

\author{Mohammed Mia$^a$ and  Fang Chen$^b$\\
$^a$ Columbia University, New York, 10027, USA\\
$^b$ McGill University, Montreal, H3A 2T8, Canada.
\vskip.07in
{\tt  mm3994@columbia.edu, fangchen@hep.physics.mcgill.ca}}

\date{.. November 2011}

\abstract{ Using the low energy limit of type IIB superstring theory, we obtain the non-extremal limit of
deformed conifold geometry which is dual to the IR limit of large N thermal QCD.
 At low
temperatures, the extremal geometry without black hole is favored while at high temperatures, the
field theory is described by non-extremal black hole geometry. We compute the ten dimensional on shell action for extremal and non-extremal
geometries and demonstrate that at a critical temperature $T_c$ there is a
 first order confinement to deconfinement phase transition. We compute $T_c$ as a function of 'tHooft coupling and study
the thermodynamics of the dual gauge theory by evaluating the free energy and entropy of the ten dimensional geometry. We find agreement with
the conformal limit while thermodynamics of non-conformal strongly coupled gauge theories is explored using the black hole geometries in
non-AdS space.     
}

\maketitle

\begin{document}

\section{Introduction}
At extreme temperatures  nuclear matter is best described by a weakly interacting gas of gluons and quarks 
due to asymptotic freedom \cite{Gross:1973id,Politzer:1973fx}. 
The weak coupling allows one to use perturbative techniques to study the thermodynamics of the system and 
at  zeroth order in perturbation theory, it is best described as a gas of free particles. But the same asymptotic  
freedom implies that the perturbative analysis must break down with decreasing temperature.  As temperature is lowered, the couplings  
get stronger and color degrees of freedom are confined. Thus nuclear matter undergoes a phase transition- from low temperature confined phase
of color neutral constituents to high temperature deconfined phase of quarks and gluons. 
However to analyze the non-perturbative confinement mechanism, 
one has to study the theory on the lattice \cite{Wilson:1974sk,Polyakov:1978vu,Susskind:1979up} or resort to effective field theories 
- 
both of which are successful along with their inherent limitations\footnote{For  analysis of high temperature phase and phase 
transitions  please consult
\cite{Gross:1980br}-\cite{Pisarski:1983ms}. For a review of recent developments in lattice QCD, please consult
\cite{Petreczky:2012rq}-\cite{Borsanyi:2012ve} and \cite{Panero:2009tv}-\cite{Panero:2012qx} for the large N limit.}.

On the other hand 'tHooft showed that when number of colors $N$ goes to infinity, only planar diagrams contribute to the amplitude
\cite{'tHooft:1973jz,Witten:1979kh}. 
Thus in the large $N$ limit, the theory drastically simplifies and one may hope that the thermodynamics could be analytically tractable. 
Furthermore, 
the structure of the planar diagrams suggest that four dimensional gauge theory maybe interpreted as a string theory \cite{Polyakov:1997tj}.
This connection becomes clear with the realization that gauge theories naturally arise from 
excitations of strings ending on branes \cite{Witten:1995im},  while on the other hand gravitons arise from 
low energy excitations of closed strings. By studying the interactions between open and closed strings, the correspondence between gauge
theory and gravity can be realized. The best studied example is the AdS/CFT correspondence proposed by Maldacena \cite{Mal-1} which essentially maps
maximally supersymmetric ${\cal N}=4$ $SU(N)$ conformal gauge theory to $AdS_5\times S^5$ geometry. In the large $N$ limit, the
 gauge theory has large 'tHooft coupling
$N\rightarrow \infty, g_s\rightarrow 0, g_sN\gg 1$ and  $AdS_5\times S^5$ geometry with small curvature $\sim 1/L^2, L\equiv
(g_sN)^{1/4}\sqrt{\alpha'}\gg 1$ has a classical description. Again the appearance of large $N$ has allowed an exact analytic description of
strongly coupled quantum gauge theory in terms of weakly coupled dual classical gravity. 

In \cite{Witt-1} an exact proposal was made to compute correlation functions of four dimensional ${\cal N}=4$ conformal $SU(N)$ gauge theory 
using partition function of
supergravity on AdS space. To study the thermodynamic properties of the gauge theory, one studies the thermodynamics of the dual geometry and for the case of CFTs,
this amounts to the study of black holes in AdS space. The AdS/CFT correspondence has been quite successful in this regard- the scaling of thermodynamic
state functions (free energy, pressure, entropy) of CFTs  with respect to temperature matches with that of black holes 
\cite{Gubser:1996de,Gubser:1998nz}. However there are no phase
transitions in conformal field theories and  
the strong coupling regime of QCD matter is far from conformal. Thus the study of
confinement to deconfinement phase transition in QCD requires careful extension of the gauge/gravity correspondence to incorporate the 
conformal anomalies. 

In principal this can be done by placing D branes in ten dimensional geometries and then studying the 
 warped geometry sourced by the fluxes and scalar fields arising from the branes. Since QCD is non supersymmetric and non-conformal, the first objective is to find gauge theories with RG flows
  arising from D brane configurations with minimal
 SUSY. There have been a lot of progress in that direction: In
 \cite{9904017, 9906194} RG flows that connected conformal fixed points at IR and UV was incorporated, 
and \cite{9909047}  connected the 
UV ${\cal N} = 4$ conformal fixed point to a ${\cal N} = 1$ confining theory. But the model with QCD like logarithmic running of the coupling 
 and 
minimal supersymmetry   
is the Klebanov-Strassler (KS)
model \cite{KS} (with an extension by Ouyang \cite{Ouyang} to incorporate fundamental matters).
Although at the highest energies the gauge theory is best described in terms of bifundamental fields 
with effective degrees of freedom
diverging, at   the lowest energies the gauge theory resembles ${\cal N}=1$ SUSY QCD. Again in the limit when effective $D3$ brane charge is
large, the gauge theory with large 'tHooft coupling has an equivalent description in terms of warped deformed cone. 

The thermodynamics of this non-conformal gauge theory is encoded in the dual geometry. A great deal of effort has been given in computing the black hole geometry dual to non-conformal thermal gauge
theories. 
For example in \cite{KT-non-ex, PandoZayas:2006sa, thorimal, Caceres:2011zn} the cascading picture of the original KS model
was extended to incorporate black-hole without any fundamental matter, while fundamental matter was accounted for in 
\cite{cotrone}. Most of the attempts are based on obtaining an effective lower dimensional action from KK reducing ten dimensional 
supergravity action. Dimensional reduction of a generic ten dimensional action  is a formidable challenge specially when there are
non-trivial fluxes and scalar fields and it is highly non-trivial to obtain a consistent truncation. Furthermore, 
it is not clear how RG flow of the dual gauge theory can be obtained since the fields in the effective action are not the dilaton or 
the 
flux in the original ten dimensional action.

On the other hand, partition function of the geometry and thus the thermodynamics of the dual gauge theory 
can be directly obtained by computing the on 
shell gravity action with appropriate boundary terms 
\cite{Gibbons-Hawking}. In a series of papers \cite{FEP}-\cite{Chen:2012me}, we proposed the ten dimensional deformed resolved conifold black
hole
geometry as the dual to UV complete gauge theory  which resembles large $N$ thermal QCD. Working with the ten dimensional geometry, we 
avoid the difficulty of KK reduction while extracting the exact RG flow of the gauge theory.  The thermal gauge theory studied in
\cite{FEP}-\cite{Chen:2012me} has a rich phase structure and as temperature is altered, we expect phase transitions.
 Our goal is to study the phase transitions at strong coupling, but to do so, we must first understand how geometries
can describe different phases. In this work, we make progress in that direction and    
 obtain the most general  ten dimensional geometry (with or without a black hole) that arises from low energy limit
of type IIB superstring theory.
Directly identifying the gauge theory partition function 
with that of the ten dimensional geometry, we
show how phase transition is realized. For any given temperature of the dual gauge theory, 
there are two geometries$-$ extremal (without black
hole) and non-extremal (with black hole) but the geometry with lower on shell action is preferred. At a critical temperature $T_c$,
 both geometries are equally likely and we have a phase transition. 
 
 Our ten dimesional solution is analytic and the corrections to the metric due to the black hole can be
 exactly written as a Taylor series in $\tilde{r}_h/r$ and $g_sM^2/N$ where $\tilde{r}_h$ is the Schwarzchild horizon and $M,N$ are number of
 five branes and effective number of three branes at some high energy. This series expansion allows us to 
 write down exact
expression for the on shell gravity action with or without the black hole and then obtain the critical horizon. However,
  in our analysis of the non-extremal geometry, we have considered constant
 axio-dilaton field. There are no D7 branes, no fundamental matter  in the  dual thermal gauge theory and no Baryochemical potential. On the
 other hand, it is straight forward to incorporate chemical potential in extremal ten dimensional geometry by considering gauge fluxes on 
 holomorphically embedded D7 branes. Note that D7 brane embedding in   
  black hole geometry is highly non-trivial in the presence of non-trivial three form fluxes\footnote{In a particular scaling limit, the D7
  embedding in a black hole geometry with considering gauge fluxes on the D7 branes along with UV completion was considered in \cite{fangmia}.
  The form of the localized sources was proposed but the exact value of the on shell action considering all the fluxes was not determined. 
  Hence the exact value of critical
  temperature was not obtained, but the scaling of the critical temperature with the number of branes should be similar to the calculation
  done in this paper.}. What we have been able to 
 obtain is the black hole geometry with only non-trivial five form $\tilde{F}_5$ and three form $G_3$ but constant dilaton and already in this simplified
 scenario, we find a first order phase transition. 
 Observe that the warped deformed cone of KS model is dual to ${\cal N}=1$ $SU({\cal M})$, ${\cal M}\gg 1$ pure gauge theory  which has no
 fundamental matter but it  confines, exhibits a 
 mass gap and has gluino condensates \cite{Sreview}. Thus the
 confinement/deconfinement phase transition we obtain should mimic the phase transition between confined glue-ball and  deconfined gluons of
 $SU(N)$ QCD without flavor. In our upcoming work \cite{Long-Keshav}, we will incorporate the effect of running dilaton field and other localized sources on the
 non-extremal geometry which will also give the dual geometric description of UV complete thermal gauge theory.         
 
 In section \ref{g/g}, we briefly discuss the supergravity equations and outline the procedure of obtaining thermodynamic state functions of
 the gauge theory. Summarizing the extremal geometry and its dual confined gauge theory in section \ref{extm}, we  propose the
 non-extremal limit of ten dimensional geometry in section \ref{Nextm}. Using the supergravity solutions for extremal and non-extremal
 geometries, in section \ref{PT}, we study transitions between the geometries which essentially describes phase transitions in the dual gauge
 theory. When the boundary of the geometry $r={\cal R}$ scales with the 'tHooft coupling $g_sN$, we find the critical temperature $T_c$ as a
 function of $g_sN$.   Finally in section \ref{Brane}, we discuss the structure of the gauge theory and propose a possible brane configuration 
 that gives rise to the supergravity solutions. Since our ultimate goal is to learn about thermal phase transitions in nuclear matter,  we
  describe the connections between the gauge theory and  large $N$ thermal QCD.

\section{Gravity Action and Gauge Theory} \label{g/g}
We start with the type IIB supergravity action including local sources in ten dimensions  \cite{DRS}\cite{GKP}:
\bg \label{Action}
S_{\rm total}&=&S_{\rm SUGRA}+S_{\rm loc}=\frac{1}{2\kappa^2_{10}}\int d^{10}x \sqrt{G}\left(R+\frac{\partial_M
\widetilde{\tau}\partial^M\bar{\widetilde{\tau}}}{2|{\rm
Im}\widetilde{\tau}|^2}-\frac{|\widetilde{F}_5|^2}{4\cdot 5!}-\frac{G_3 \cdot \bar{G}_3}{12 {\rm Im}\widetilde{\tau}}\right)\nonumber\\
&+&
\int_{\Sigma_8} C_4\wedge R_{(2)}\wedge R_{(2)}+\frac{1}{8i\kappa_{10}^2}\int \frac{C_4\wedge G_3\wedge \bar{G}_3}{{\rm Im} \widetilde{\tau}}+S_{\rm loc}
\nd
in Einstein frame, where $\widetilde{\tau}=C_0+i e^{-\phi}$ is the axio-dilaton with $C_0$ being the axion and $\phi$ the dilaton field and
$\widetilde{F}_5=dC_4-\frac{1}{2}C_2\wedge H_3+\frac{1}{2} B_2\wedge F_3$ is the five-form flux sourced by the D3 or fractional D3 branes. Here $G_3\equiv F_3-\widetilde{\tau}
H_3$ with $F_3$ the RR three form flux and $H_3=dB_2$  the NS-NS three form flux with $B_2$ being the
NS-NS two form while $C_2$ is the RR two form. We also have
$C_4$ the four-form potential,
$G=\sqrt{-{\rm det}~g_{MN}},M,N=0,..,9$ with $g_{MN}$
being the metric, $R_{(2)}$  the curvature two-form, and $S_{\rm loc}$ is the action for localized sources in the system
(i.e. D branes).

The above action (\ref{Action}) is the most general
supergravity action obtained from type IIB superstring action with fluxes and localized sources. Minimizing the action leads to the 
following Einstein equations
\bg \label{ricci_T}
R_{\mu\nu}&=&-g_{\mu\nu} \left[\frac{G_3 \cdot \bar{G_3}}{48\; {\rm
Im}\widetilde{\tau}}+\frac{\widetilde{F}_5^2}{8\cdot 5!}\right]+\frac{\widetilde{F}_{\mu
abcd}\widetilde{F}_\nu^{\;abcd}}{4 \cdot 4!}
+\kappa_{10}^2 \left(T_{\mu\nu}^{\rm loc}-\frac{1}{8} g_{\mu\nu} T^{\rm loc}\right)\nonumber\\
R_{mn}&=&-g_{mn} \left[\frac{G_3 \cdot \bar{G_3}}{48 \;{\rm
Im}\widetilde{\tau}}+\frac{\widetilde{F}_5^2}{8\cdot 5!}\right]+\frac{\widetilde{F}_{m
abcd}\widetilde{F}_n^{\;abcd}}{4 \cdot 4!}+\frac{G_m^{\;bc}\bar{G}_{nbc}}{4\;{\rm Im}\widetilde{\tau}}
+\frac{\partial_m \widetilde{\tau} \partial_n \widetilde{\tau}}{2\;|{\rm Im}\widetilde{\tau}|^2}\nonumber\\
&+&\kappa_{10}^2 \left(T_{mn}^{\rm loc}-\frac{1}{8} g_{mn} T^{\rm
loc}\right) \nd
where $\mu,\nu=0,..,3$; $m,n=4,..,9$ and we have assumed that the fluxes and axio-dilaton
only depends on coordinates $x^m$.

The equation of motion for $G_3$ can be expressed in terms of a
seven-form $\Lambda_7=\ast_{10}G_3-iC_4\wedge G_3$ in the following
way,
\begin{eqnarray}\label{g3eom}
d\Lambda_7-\frac{i}{\textrm{Im}\widetilde{\tau}}d\widetilde{\tau}\wedge\textrm{Re}\Lambda_7=0
\end{eqnarray}
where typically $\Lambda_7$ quantifies  the deviations from the imaginary self dual (ISD)
behavior.
Minimizing the action (\ref{Action}) also gives the Bianchi identity for the five-form flux \cite{GKP}
\begin{equation}\label{bianchi}
d\widetilde{F}_5=-\frac{G_3\wedge \bar{G_3}}{2i\textrm{Im}\widetilde{\tau}}+ 2\kappa_{10}^2\; T_3\; \rho_3^{\rm loc}
\end{equation}
where $\rho_3^{\rm loc}$ is the D3 charge density from the localized sources and $T_3$ is the brane tension. In the presence of D3 branes,  $\rho_3^{\rm loc}$ is a
delta function peaked at the location of the branes while D7 or fractional D5 branes may also contribute to D3 charge.

To keep lorentz invariance along the space-time direction we assume the
self-dual five-form has the form
\begin{eqnarray}\label{dF5}
\tilde{F}_5=(1+\ast)d\alpha \wedge dt\wedge dx\wedge dy\wedge dz.
\end{eqnarray}
where $\alpha(x^m)$ is a scalar field, function of the internal coordinates $x^m$.
We take a general metric ansatz:
\begin{eqnarray}\label{metric}
ds^2&=&-e^{2A+2B}dt^2+e^{2A}(dx^2+dy^2+dz^2)+e^{-2A-2B}\tilde{g}_{mn} dx^m dx^n\nonumber\\
\tilde{g}_{mn} dx^m dx^n&=&\Big(a(r)dr+k(r)ds^2_{\mathcal {M}_5}\Big)
\end{eqnarray}
where $B\neq 0$ characterizes the existence of a black hole. Using this metric ansatz we can express
$\Lambda_7$ as \bg\label{lambdaeqon} \Lambda_7 ~ = ~
\left[e^{4A+B}\ast_{6}G_3-i\alpha G_3\right] \wedge dt\wedge
dx\wedge dy\wedge dz \nd The above choice of $\Lambda_7$ leads us to
three different classes of solutions from the $G_3$ EOM
\eqref{g3eom}. These three classes can be tabulated in the following
way: \vskip.1in

\noindent $\bullet$ If $\alpha=e^{4A+B}$ in \eqref{lambdaeqon} and
$\Lambda_7 = d\Lambda_7 = 0$ then $G_3$ must be imaginary self dual (ISD). When $B=0$ then
this is the same as GKP solution \cite{GKP}, and in this
case $\widetilde{\tau}$ is not restricted and
\begin{eqnarray}
\tilde{R}_{mn}=\frac{\partial_m\widetilde{\tau}
\partial_n\bar{\widetilde{\tau}}}{2\mid\textrm{Im}\widetilde{\tau}\mid^2}-3\widetilde{\triangledown}_m\partial_n B -\frac{3}{2} \partial_mB\partial_nB
\end{eqnarray}
where $\tilde{R}_{mn}$ is the Ricci tensor for cone metric $\tilde{g}_{mn}$.

\noindent $\bullet$ If $\alpha\neq e^{4A+B}$ then we can take
$\Lambda_7\neq 0$ but keep $d\Lambda_7=0$ and $d\widetilde{\tau}=0$. This means
$\widetilde{\tau}$ is a constant\footnote{Or $\widetilde{\tau} = d\lambda_{-1}$ i.e $d$ of a
($-1$)-form. The functional form for the ($-1$)-form is non-trivial,
so this option is more cumbersome to use.}.

\noindent $\bullet$ If $\alpha\neq e^{4A+B}$ then we can again take
$\Lambda_7 \ne 0$ but now $d\Lambda_7\neq 0$ and $d\widetilde{\tau}\neq 0$ such
that \eqref{g3eom} is satisfied. This means both axion and the
dilaton could run in this scenario.

\vskip.1in

The first possibility is studied in the appendix \ref{apdx}. We are more interested in the last two options because
it is well established that AdS black hole solution corresponds to conformal $SU(N)$ gauge theory, and non-AdS
black hole solutions should reduce to AdS black hole solutions in certain limit. It is obvious that in the first
case at $G_3=0$ it does not recover AdS black hole solution while the last two cases obviously do.

Suppose now we find a solution to the above set of Einstein
equations (\ref{ricci_T}) along with the flux equations
(\ref{g3eom}), (\ref{bianchi}). The resulting geometry with the metric (\ref{metric}) has the topology of $Y^5\times {\cal S}^5$ and
in particular for $G_3=0$, $Y^5$ can be a five dimensional AdS space with or without black holes which has throat radius $L$ while ${\cal S}^5$ can be a compact five sphere $S^5$ with radius
$L$.
For non-vanishing three form flux $G_3$, $Y^5$ will be a non-AdS space and the size of ${\cal S}^5$ given by $e^{-2A-2B} k(r)$ will diverge for
large $r$. However for fixed radial coordinate $r$, the geometry has topology $M_4\times {\cal S}^5$ where $M_4$ is four dimensional Minkowski
space and ${\cal S}^5$ has fixed radius. This way at fixed radial location (which we interpret as the boundary),
we can obtain a four dimensional
manifold $M_4$ by integrating over the compact space ${\cal S}^5$. The gauge theory is defined on Minkowski space
$M_4$ while the manifold $Y^5\times {\cal S}^5$ is the dual geometry for the gauge theory.

According to AdS/CFT correspondence, the partition function of
string theory on $AdS_5\times S^5$ should coincide with the partition function of $\mathcal{N}=4$ super-
Yang-Mills theory. For generalized gauge/gravity correspondence, we can identify the gauge theory partition function with that of ten
dimensional dual geometry,
\begin{eqnarray}\label{KS16}
\mathcal{Z}_{\rm gauge}&=&e^{-F/T}=\mathcal{Z}_{\rm gravity}\simeq e^{-S^{\rm ren}_{\rm gravity}}\nonumber\\
S^{\rm ren}_{\rm gravity}&=&S_{\rm total} +S_{GH}+ S_{\rm counter}
\end{eqnarray}
where $F, T$ are  free energy and temperature of the gauge theory, $S_{\rm total}$ is given by (\ref{Action}), $S_{GH}$ is the Gibbons-Hawking boundary term \cite{Gibbons-Hawking}
and $S_{\rm counter}$ is the counter term necessary to renormalize the action. The gravity action is evaluated on $Y^5\times {\cal S}^5$ 
and then wick rotated $t=i \tau$ to obtain the Euclidean on shell value. The $\simeq$ is because we have ignored
all the $\alpha'$ corrections and loop corrections.

 Using (\ref{KS16}), we can in principle obtain all the thermodynamic quantities for the four dimensional gauge theory by considering the ten
 dimensional action $S^{\rm ren}_{\rm gravity}$ on shell. For instance, free energy $F$ and internal
energy $E$ of the gauge
theory are given by
\bg \label{FreeE}
F&=&- \rm{T} ~\rm{log} \left({\cal Z}_{\rm gauge}\right)
=-T~\rm{log}\left( {\cal Z}_{\rm gravity}\right)=T ~S^{\rm ren}_{\rm gravity}\nonumber\\
E&=&T^2\frac{\partial}{\partial T} \left({\rm log} {\cal Z}_{\rm gauge}\right)=\frac{\partial S^{\rm ren}_{\rm gravity}}{\partial \beta }
\nd
where $T$ is the temperature. Knowing the free energy one gets the pressure $p$ and entropy $s$
\bg  \label{pE}
p&=&-\left(\frac{\partial F}{\partial V_3}\right)_T\nonumber\\
s&=& -\left(\frac{\partial F}{\partial T}\right)_{V_3} \nd where $V_3=\int
d^3x$ is the volume of three dimensional flat space.

 Our primary concern for this paper is to obtain the ten dimensional geometries that can arise from the action (\ref{Action}). However, the same action can give rise to
more than one manifolds. In fact, just like the case for AdS space
discussed by Hawking and Page \cite{Hawking:1982dh} and elaborated
by Witten \cite{Witten:1998zw}, there are two manifolds
$X^1=S^1\times M^1$ and $X^2=S^1\times M^2$ that minimizes the
action (\ref{Action}). The manifold which has lower value for the on shell action for a given temperature of the dual gauge theory
will be preferred. Since $X^1$ and $X^2$ are distinct geometries, the thermodynamics of the gauge theory will be different at different
temperatures, depending on which geometry is preferred. This means $X^1$ and $X^2$ will correspond to different phases of the gauge theory and
we will now analyze the manifolds in some details
bellow.

\subsection{Extremal geometry and confinement}\label{extm}
 The metric of the extremal geometry $X^1$ without any black hole, i.e. $B=0$, is given by \cite{KS} (with Minkowski signature)
\bg\label{metd}
ds^2 &=& e^{2A}
\Big[-dt^2+dx^2+dy^2+dz^2\Big] +e^{-2A}\bar{g}_{mn} dx^m dx^n
\nd
where $e^{-4A}=h(\rho)$ and $\bar{g}_{mn}$ is the metric of the deformed cone
\bg\label{inmateda}
&&\bar{g}_{mn} dx^m dx^n ~ = \frac{1}{2}{\cal A}^{4/3} K(\rho)\Big[\frac{1}{3K^3(\rho)}\left(d\rho^2+(g^5)^2\right)+{\rm
cosh}^2\left(\frac{\rho}{2}\right)\left[(g^3)^2+(g^4)^2\right]\nonumber\\
&&+ {\rm
sinh}^2\left(\frac{\rho}{2}\right)\left[(g^1)^2+(g^2)^2\right]\Big]
\nd
where ${\cal A}$ is a constant, $g^i,i=1,..,5$ are one forms given by

\bg\label{oneforms}
&&g^1=\frac{e^1-e^3}{\sqrt{2}},~~~~g^2=\frac{e^2-e^4}{\sqrt{2}}\nonumber\\
&&g^3=\frac{e^1+e^3}{\sqrt{2}},~~~~g^4=\frac{e^2+e^4}{\sqrt{2}},~~~g^5=e^5\nonumber\\
&& e^1\equiv-{\rm sin}\theta_1 \;d\phi_1, ~~~~ e^2\equiv d\theta_1\nonumber\\
&& e^3\equiv {\rm cos}\psi \;{\rm sin}\theta_2 \;d\phi_2-{\rm sin}\psi \;d\theta_2,\nonumber\\
&& e^4\equiv {\rm sin}\psi \;{\rm sin}\theta_2\; d\phi_2+{\rm cos}\psi\; d\theta_2,\nonumber\\
&& e^5\equiv d\psi +{\rm cos}\theta_1\; d\phi_1+{\rm cos}\theta_2
\;d\phi_2 \nd and \bg K(\rho)=\frac{\left({\rm
sinh}(2\rho)-2\rho\right)^{1/3}}{2^{1/3}{\rm sinh}\rho}. \nd The
 three form fluxes $G_3$ on the deformed
cone is Imaginary Self Dual (ISD) to preserve the $\mathcal{N}=1$
supersymmetry
 while the five form fluxes is self dual \bg \label{F_5} ~~\ast_{6} G_3=i\; G_3,~
\widetilde{F}_5=(1+\ast) dh^{-1}\wedge dt\wedge dx\wedge dy\wedge
dz=dC_4+B_2\wedge F_3. \nd

Note that the metric (\ref{inmateda}) is Ricci flat, that is the Ricci tensor for the metric $\bar{g}_{mn}$ denoted by $\bar{R}_{mn}=0$
 (but Ricci tensor for the metric $g_{mn}$ denoted by $R_{mn}\neq 0$), while minimizing the action (\ref{Action}) with non-zero
$\partial \widetilde{\tau}$ and localized sources will give rise to an internal metric which is not Ricci flat in general. In particular, the Ricci
tensor for the warped metric is
\bg \label{R_fangA}
R_{\mu\nu}&=&\eta_{\mu\nu}
\frac{1}{4h(\rho)}\bar{\triangledown}^2{\rm log}h(\rho)\nonumber\\
R_{mn}&=&\bar{R}_{mn} -\frac{\bar{g}_{mn}}{4} \tilde{\triangledown}^2{\rm log}h(\rho)
-\frac{1}{2}\partial_m{\rm log} h(\rho)\partial_n{\rm log}h(\rho)
\nd
where
\bg\label{Laplacian}
\bar{\triangledown}^2=\bar{g}^{mn}\partial_m\partial_n
+\partial_m\bar{g}^{mn}\partial_n+\frac{1}{2}\bar{g}^{mn}\bar{g}^{pq}\partial_n\bar{g}_{pq}
\partial_m
\nd
 Now using (\ref{R_fangA}) in (\ref{ricci_T}) with the metric given by (\ref{metd}), one readily gets that $\bar{R}_{mn}\sim
 \partial_m\widetilde{\tau}\partial_m\widetilde{\tau}+ {\rm local \;sources}$.
 However, F-theory \cite{vafaF,Sen:1996vd} dictates that the running of the
axio-dilation field is $\partial\widetilde{\tau}\sim {\cal O}(g_sN_f)$ where $N_f$ is the number of seven branes
and we will only consider terms which are of ${\cal
O}(g_sN_f)$ ignoring all higher order terms. If we also ignore the localized sources, then indeed we obtain $\bar{R}_{mn}=0$ \cite{fangmia,GKP}.

Even though $\bar{g}_{mn}$ remains the same the warp factor is
different in different regions. In the IR region we assume the D7
branes are far away and thus the axion-dilaton is constant. The warp
factor, $B_2$ and $F_3$ in this region  are given in \cite{KS} and
calculated to the first order. The warp factor can be written in the
following form
\begin{eqnarray}\label{warpf1}
h(\rho)=c_i\rho^i
\end{eqnarray}
where the coefficient $c_i$ can be treated as constants. When doing
so it is assumed that
\begin{eqnarray}
\ast dh^{-1}\wedge dt\wedge dx^1\wedge dx^2\wedge dx^3=B_2\wedge F_3
\end{eqnarray}
This is correct if the number of D3 branes is multiple of that of
the D5 branes. In the rest of the paper we will restrict to this
special case\footnote{In general \begin{eqnarray} 
\ast
dh^{-1}\wedge dt\wedge dx^1\wedge dx^2\wedge dx^3=B_2\wedge F_3+Ng^1\wedge
g^2\wedge g^3\wedge g^4\wedge g^5,
\end{eqnarray} where $N$ is the D3 brane charge at certain energy scale $r_0$ where $B_2\wedge F_3=0$.
If $N$ is written as $N=lM+k$ with $0\leq k<M$, then $lM$ can be
absorbed into $B_2\wedge F_3$ leaving only the $k$ term. This is the
cascading shown in the gravity side. }.

The warped deformed conifold with $h(\rho)$ given by (\ref{warpf1})
for small $\rho$ as proposed in \cite{KS} in fact removes the IR
singularity of the warped conifold \cite{KT} and gives rise to
linear confinement in the dual gauge theory. Unlike the regular
cone, the deformed cone has a blown up  $S^3$ at the tip of the cone
and this finite size of the $S^3$ removes the IR divergence of the
fluxes \cite{KS}. Observe that  near $\rho= 0$, the metric in
(\ref{inmated}) reduces to that of $S^3$ \bg\label{inmated0}
\bar{g}_{mn} dx^m dx^n \sim {\cal A}^{4/3}(2/3)^{1/3}
\left[\frac{1}{2}(g^5)^2+(g^3)^2+(g^4)^2\right] \nd
 which implies that the radius of $S^3$ at the tip of the cone is of ${\cal O}({\cal A}^{2/3})$. The finite size of $S^3$ at $\rho=0$ gives finite
 value for the three form flux strength $|F_3|^2,|H_3|^2$ at the tip of the cone. This way, the IR singularity of the fluxes are removed, which consequently
 removes the IR singularity of the warp factor.

 On the other hand, this constant ${\cal A}$ modifies the embedding equation for the cone and is
  related to the expectation values of gauge invariant operators\footnote{At the bottom of the cascade when {\it all} D3 branes disappear and
  only fractional branes remain, we have pure glue SU(M) with gluino condensates $\langle \lambda \lambda \rangle$ \cite{Sreview}. The dual
  geometry partially captures the feature of the pure glue theory and we expect  non-zero
  ${\cal A}$ corresponds to non-zero $\langle \lambda \lambda \rangle$. } in the
 dual gauge theory \cite{KS}. Hence a non-zero $\mathcal{A}$ results in non-zero expectation value and
  corresponds to spontaneous  breaking of $Z_{2M}$ group down to $Z_2$ where $M$ is the
 number of fractional three branes placed at the tip of the cone. Note that in the absence of  D7 branes, this
 spontaneous broken symmetry is not identical to the chiral symmetry of QCD as there are no fundamental flavor. But even in the presence of
 seven branes, the deformed cone metric (\ref{metd}) with warp factor (\ref{warpf1}) will be valid by considering the D7 branes as probes. The
 probe seven branes will give rise to fundamental matter and depending on their embedding, can lead to breaking of flavor chiral symmetry
 just like in QCD \cite{Dymarsky:2009cm}.

Coming back to the metric (\ref{inmateda}), observe that with a
change of coordinates \bg \label{rrho} r^3={\cal A}^2 e^\rho \nd for large
$\rho$, the metric becomes \bg\label{inmate1} &&\bar{g}_{mn} dx^m
dx^n\sim dr^2+r^2\left(\frac{1}{9} (g^5)^2+\frac{1}{6}\sum_{i=1}^{4}
(g^i)^2\right) \nd which is the metric of regular cone with base
$T^{1,1}$. Thus only for small radial coordinate $\rho$, the
internal metric is a deformed cone while at large $\rho$, we really
have a regular cone with topology of $R\times T^{1,1}$.

However for large $\rho$, we can no longer ignore
the running of the $\widetilde{\tau}$ field as we will be near the seven branes. As we have a regular cone for large $\rho$, we can use
 Ouyang's holomorphic embedding of seven branes \cite{Ouyang} to determine the running of the $\widetilde{\tau}$ field along with the modified flux
 $G_3$ (which is again ISD) and $\tilde{F}_5$. Then for $\rho> 0$, using change of coordinates (\ref{rrho}), we obtain the warp factor
 \bg \label{h(r)}
 &&h(r,\theta_1,\theta_2)=\frac{\alpha'^2}{r^4}\Bigg(\alpha_0+\frac{81g_sM^2}{8} \;{\rm log}\left(\frac{r}{r_l}\right)
 \left[1+\frac{3g_s N_f}{2\pi}\left({\rm log}\left(\frac{r}{r_{\rm min}}\right)
+\frac{1}{2}\right)\right]\nonumber\\
&+&\frac{81g_s^2 M^2 N_f}{32\pi}\;{\rm log}\left(\frac{r}{r_{\rm min}}\right)\; {\rm
log}\left(\rm{sin}\frac{\theta_1}{2}\rm{sin}\frac{\theta_2}{2}\right)\Bigg)
\nd
 where $N_f$ is
 the number of D7 branes which is present in the action (\ref{Action}) and source the $\widetilde{\tau}$ field while $r_{\rm min}$
 is the minimum radial distance reached by the seven brane.
 The exact brane configuration that could give rise to such a warp factor in the dual geometry will be discussed in section \ref{Brane}. Also
  $\alpha_0,r_l$ are  constants which will be determined by
 matching the above solution for large $\rho$ with the solution (\ref{warpf1}) valid
 at small $\rho$.  In fact $\alpha_0$ is proportional to the D3
 brane charge at $r=r_l $. To see this we notice that
 on the gravity action, we do not have any branes - only fluxes.
 However,
 due to the duality, the units of the fluxes on the gravity side
 should equal to the number of branes on the gauge theory side. This way
 the units of five form fluxes can be identified as the effective charge:
 \bg\label{Neff}
 N_{\rm eff}&=&\frac{1}{2 \kappa_{10}^2T_3}\int_{T^{1,1}} \tilde{F}_5
 =-\frac{1}{216 \kappa_{10}^2 T_3}\int r^5\frac{dh}{dr}\; g^1\wedge g^2\wedge g^3 \wedge g^4 \wedge g^5
 \nd
  where $T_3=\mu_3=(2\pi)^{-3}\alpha'^{-2}$  is the three brane tension and the Gaussian surface  is  
  warped $T^{1,1}$ with radius $\sqrt{r^2 \sqrt{h}}$  that
  encloses Minkowski space.
  Using (\ref{h(r)}) in
 (\ref{F_5}),  we readily get from (\ref{Neff}) that $N_{\rm eff}$ depends on $r$. In particular at $r=r_l$
 \bg
 N_{\rm eff}(r=r_l)&=& \frac{ V_5 \alpha_0\alpha'^2}{54 \kappa_{10}^2 T_3} + {\cal O}(g_sM^2, g_s^2M^2N_f)\nonumber\\
 V_5&\equiv& \int d\psi d\phi_1 d\phi_2 d\theta_1 d\theta_2 \; \rm{sin}\theta_1 \rm{sin} \theta_2
 \nd
 
   Using the exact same argument, we  get $N_{\rm
 eff}(\rho=0)=0$ as $\partial h/\partial\rho=0$ at $\rho= 0$ \cite{KS}. This implies that there are no effective D3 branes, they have cascaded away at the IR
 and we are left with only fractional D3 branes.

 Note that the holomorphic D7 brane embedding of Ouyang \cite{Ouyang} only gives $U(N_f)$ flavor
 symmetry group. However by considering $N_f$ number of $D7$ and $\bar{D}7$ branes, the chiral $U_L(N_f)\times U_R(N_f)$ symmetry can be
 realized \cite{Dymarsky:2009cm,Chen:2012me} but then the warp factor (\ref{h(r)}) will also be modified.  In addition, the warp factor
 (\ref{h(r)}) only makes sense in the large $\rho$ region which means $r_{\rm min}$ is large. Thus the fundamental matter fields resulting from the holomorphic
 embedding have large mass and we do not expect chiral symmetry to hold for massive flavors. As supersymmetry is not broken at zero
 temperature for the Ouyang embedding, the warp factor (\ref{h(r)}) corresponds to a gauge theory with massive fundamental matter that is supersymmetric at zero
 temperature. Recently in \cite{Dymarsky:2011ve}, the dual to  non-supersymmetric confining gauge theory was proposed and the metric (\ref{metd}) is
 consistent with this proposal with the understanding that $\bar{g}_{mn}$ in (\ref{inmated}) be replaced with the perturbed metric $\bar{g}_{mn}^1$ which is also
 Ricci flat.
 We will analyze  possible brane configurations that may give rise to SUSY and non-SUSY gauge theories with fundamental representation matter
 in section \ref{Brane}.    .

 The solution (\ref{h(r)}) leads to divergence of $N_{\rm eff}$
  for very large $r$ i.e very large energies. But this is expected as the dual
 theory is a cascading gauge theory with effective color diverging in the UV
 \cite{KS, Sreview,Benini:2011cm}. However our goal is to study thermal
 QCD which becomes conformal and  free in asymptotically large energies. Thus the dual geometry with warp factor
 (\ref{h(r)}) cannot be relevant for a QCD like theory and must be modified for $r\sim \Lambda \rightarrow \infty$. 
 In   \cite{Mia:2010tc,Chen:2012me} a proposal was
 made to modify the UV dynamics of the cascading gauge theory by introducing anti five branes separated from $M$ D5 and $N$ D3 branes along
 with embedding $D7-\bar{D}7$ branes to account for fundamental matter and chiral symmetry breaking. 
 The additional 
 branes modify $G_3$ and $\tilde{F}_5$ while the form of the $\widetilde{\tau}$ can be obtained from F-theory \cite{vafaF,Sen:1996vd}. 
 Note that the $D5-\bar{D}5$, $D7-\bar{D}7$ system behaves as dipoles and thus away from branes, their contribution to flux and axio-dilaton
 is not divergent and behaves as $1/r^i, i>0$. One can consider brane embeddings such that the fluxes and axio-dilaton fields in the dual
 geometry is well defined everywhere in the bulk geometry  
 while the form of the metric
 (\ref{metd}) remains unchanged. However, the internal metric does not remain Ricci flat anymore and the metric (\ref{inmate1}) gets corrections of
 ${\cal O}(g_s^2N_f^2)$.

 The fluxes proposed in \cite{Mia:2010tc} essentially give $B_2\rightarrow 0$ for $r\gg r_0$ \footnote{$r_0\sim \Lambda_0$ is 
 related to the higgs mass
  in
the dual gauge theory. For details of the higgising mechanism, please consult \cite{Chen:2012me} } while the axio dilaton field behaves
 as $\widetilde{\tau}\sim b_i/r^i,b_i\sim {\cal O}(g_s N_f)$     for large $r$. Thus for asymptotically large distances, we end up with constant
 axio-dilaton field and a vanishing NS-NS two form $B_2$ and the geometry behaves as an $AdS_5\times T^{1,1}$. This means that addition of
 anti five branes have slowed down the cascade and the theory has no Landau poles and the associated  UV divergences. The warp factor becomes

 \bg \label{h_UV}
 h(r,\Theta)=\frac{1}{r^4}\sum_i\left(\frac{a_i(\Theta)}{r^i}\right)
 \nd
  for large $r$ where $\Theta\equiv (\theta_i,\phi_i,\psi)$. Using (\ref{Neff}) one readily gets that $N_{\rm eff}(r=\infty)\sim a_0$ and hence
  does not diverge. As $N_{\rm eff}$ reaches a fixed value with the geometry becoming $AdS_5\times T^{1,1}$,
  the gauge theory reaches a conformal fixed point. However, for supergravity approximation to be valid, we need $N_{\rm eff}(\infty)\sim
  a_0\gg 1$ which means the 'tHooft coupling  $g_sN_{\rm eff}(\infty)$ is still very large .
  Although we have a conformal field theory at large
  scale $\Lambda=r\rightarrow \infty$, the gauge group has large number of effective colors with large 'tHooft coupling and thus the gauge
  theory is not asymptotically free. However, the Yang Mills coupling of our gauge theory $(g_1,g_2)$ behaves the
  following way

  \bg
  \frac{1}{g_1^2}+\frac{1}{g_2^2}&=&e^{-\phi}\nonumber\\
  \frac{1}{g_1^2}-\frac{1}{g_2^2}&=&e^{-\phi}\int_{S^2} B_2
  \nd
At asymptotically large energies, that is $r\rightarrow \infty $, $B_2\rightarrow 0$ and we have $g_1\simeq g_2=g_{YM}$ with
\bg
\lim_{\Lambda\rightarrow \infty}\frac{2}{g_{YM}^2}=\frac{1}{g_s}
\nd
Now we can take the limit $g_s\rightarrow 0$ by keeping $g_s N_{\rm eff}$ fixed, which means $g_{YM}\rightarrow 0$ at asymptotically
large energies, just like in QCD. However, since  $g_s N_{\rm eff}$ is held fixed at large values, we have large 'tHooft coupling while for
QCD, 'tHooft coupling goes to zero. 

  In summary, the extremal geometry  takes the simple form given by (\ref{metd}) while the warp factor is
  given by (\ref{warpf1},\ref{h(r)}) and (\ref{h_UV}) for various range of the radial coordinate $\rho$. The fluxes for various 
  regions of the geometry can be found in \cite{Mia:2010tc}.

  Now to obtain thermodynamic state functions such as free energy, entropy and pressure of the gauge theory that is dual to the geometry
  $X^1=Y^5\times {\cal S}^5$, we
  must obtain the on shell supergravity action on the manifold $X^1$. However note that,
   since there is no black hole horizon in the geometry $X^1$, the Euclidean renormalized on shell action $S^{\rm ren}_{\rm gravity}=\beta F$
     is independent of horizon with
    $F$
   independent of $T$. Then using (\ref{FreeE}) one obtains that
   the free energy is independent of temperature and (\ref{pE}) readily gives
  \bg \label{s0}
  s=-\frac{\partial F}{\partial T}=0
  \nd

The above result for entropy is consistent with the confined phase of large $N$ gauge theory. For confined phase, $Z_N$ symmetry is preserved and
physical quantities do not depend on the temporal radius at ${\cal O}(N^2)$, due to large $N$ volume independence \cite{Eguchi:1982nm},\cite{Gocksch:1982en}. 
This means free energy is independent of temperature and entropy is zero at ${\cal O}(N^2)$.  
  Thus the gauge theory  dual to the geometry $X^1$ is in the confined phase.
  We will later see that for $X^2$, entropy is nonzero- indicating we
  have a deconfined phase and degrees of freedom have been released at high temperature.


\subsection{Non extremal geometry}\label{Nextm}

The non-extremal geometry $X^2$ is a regular  black hole
 with $B\neq 0$ in (\ref{metric}). The internal manifold is typically
 a resolved-deformed conifold that mimics the
symmetries (or breaking of symmetries) of the dual gauge theory.
 For simplicity, we will consider constant axio-dilaton field, that is there are no seven branes in the dual gauge theory, i.e  $N_f=0$. We make the following ansatz for
 the internal metric
 \bg\label{inmated}
 \tilde{g}_{mn}
dx^m dx^n ~ &=& a(r)dr^2+
k(r)\tilde{g}_{pq}^{(1)} dx^p dx^q\nonumber\\
\tilde{g}_{pq}^{(1)} dx^p
dx^q&\equiv&\Big[\frac{1}{9}(g^5)^2+\frac{1}{6}\Big(\sum_{i=1}^4
(g^i)^2\Big)\Big]+\tilde{g'}_{pq}dx^p dx^q \nd where $g^i$ are
defined in (\ref{oneforms}), $k(r)=r^2e^{2B}b(r)$ and $x^p,x^q$
represent angular coordinates. Note that when there are no black
holes i.e. horizon $r_h=0$  and no three form flux i.e. $M=0$, we
have $\tilde{g'}_{pq}=0$ and $a(r)=b(r)=e^{2B}=1$. On the other
hand, as we will see explicitly, fluxes enter the Einstein equations
with terms of ${\cal O}(g_sM^2/N)$, where 
\bg
N\equiv \alpha_0\simeq \frac{54 \kappa_{10}^2 T_3 N_{\rm eff}(r=r_l)}{V_5\alpha'^2}
\nd is proportional effective
number of D3 branes at certain energy scale $r_l$. These scalings with horizon and flux strength
indicate that at the lowest order, we have \bg \label{g'}
\tilde{g'}_{pq}&=&{\cal O}(r_hg_sM^2/N)\nonumber\\
a(r)&\equiv&1+ a_1(r)= 1+{\cal O}(r_hg_sM^2/N), ~~ b(r)\equiv 1+ b_1(r)= 1+{\cal O}(r_hg_sM^2/N)\nonumber\\
\nd
Note that with $\tilde{g'}_{pq}\neq 0$  the two $S^2$ are
squashed and topologically the base of the cone  is no longer $S^2\times S^3$.

Before solving the equations of motions, we would like to
discuss some subtleties:

\noindent $\bullet$ We are studying the gravity dual of
the gauge theory and these gravity solutions are generated by
fluxes only. The gauge theory lives on N D3 branes and M D5 branes placed at the tip of a regular conifold, the back reactions of the branes 
will give us a warped geometry.
 There are fluxes in these gravity solutions whose integral $\int F$ is non-vanishing, however that does not mean there is a $\delta$ function or brane source in such solutions. Actually $\int F$  can be non-zero without any source if the volume of integrated space is non-vanishing anywhere. This is what happens in $AdS_5\times S^5$ case, $\int_{S^5} F_5\neq 0$ where $S^5$ is a 5D sphere with constant radius.  Particularly in  Klebanov-Tseytlin solution, the radius of the warped $T^{1,1}$ becomes zero at non-zero $r$ and becomes
ill defined at very small $r$, leading to singularities. This singularity was resolved by  Klebanov and Strassler \cite{KS}
 who pointed out that
 at low energies,
the gauge theory superpotential receives quantum corrections \cite{Affleck:1983mk} and the corresponding dual geometry is no longer the
warped regular cone but it must be warped  deformed cone whose $S^3$ is non-vanishing at $r=0$. Thus even
though $\int_{T^{1,1}}F_5\neq 0$ and $\int_{S^3}F_{3}\neq 0$ there
is no $\delta$ function in either flux equations or Einstein
equations. In section (\ref{Brane}), we will outline the brane configuration
that may give rise to the dual gravity solutions that we present
here.

\noindent $\bullet$
The Schwarzschild black hole is a vacuum solution of Einstein equations
with the stress energy tensor $T_{mn}=0$. The horizon radius is an independent parameter and
not determined by the Einstein equations. By considering the weak field limit, one can obtain the geometric
mass of the black hole from the horizon radius which is a free parameter. Similarly by placing D3 branes at the tip of regular cone, we can generate AdS
Schwarzschild black holes and again the black hole mass is a free parameter independent of the number of D3 branes.
 In our solution
discussed below we will find a charged black hole and we interpret the charge and mass arising from the D branes placed
at the tip of the regular cone. We will see that the black hole is {\it not a Schwarzschild black hole} and the horizon is no longer independent of the
number of D branes. However, the solution can be built from the Schwarzschild solution and thus the vacuum solution plays a crucial role in
our analysis.

Now lets look at the equations of motions. We first consider the
external components of the Einstein equations in (\ref{ricci_T}).
Using the form of $\tilde{F}_5$ (\ref{dF5}), the first equation in
(\ref{ricci_T}) becomes \bg\label{Ricci_Min}
R_{\mu\nu}&=&-g_{\mu\nu} \left[\frac{G_3 \cdot \bar{G_3}}{48\; {\rm
Im}\widetilde{\tau}}+\frac{e^{-8A-2B}
\partial_m\alpha \partial^m\alpha}{4}\right]+\kappa_{10}^2
\left(T_{\mu\nu}^{\rm loc}-\frac{1}{8} g_{\mu\nu} T^{\rm loc}\right)
\nd
On the other hand, the Ricci tensor in the Minkowski direction takes
the following simple form
\bg \label{ricci_Min}
R_{\mu\nu}&=&-\frac{1}{2} \left[\partial_m (g^{mn} \partial_n
g_{\mu\nu})+ g^{mn} \Gamma^M_{nM}\partial_m g_{\mu\nu}-
g^{mn}g^{\nu'\mu'}\partial_m g_{\mu'\mu}
\partial_n g_{\nu'\nu}\right]
\nd
where $\nu',\mu'=0,..,3$ and $\Gamma^M_{nM}$ is the Christoffel
symbol. Now using the ansatz (\ref{metric}) for the metric,
(\ref{ricci_Min}) can be written as

\bg\label{ricci_Min_a}
R_{tt}&=&e^{4(A+B)}\left[\widetilde{\triangledown}^2(A+B)-3\widetilde{g}^{mn}\partial_nB\partial_m(A+B)\right]\nonumber\\
R_{ij}&=&-\eta_{ij}
e^{2(2A+B)}\left[\widetilde{\triangledown}^2A-3\widetilde{g}^{mn}\partial_nB\partial_mA\right]
\nd where the Laplacian is defined in (\ref{Laplacian}).

The set of equations can be simplified by taking the trace of the first equation in (\ref{ricci_T}) and using
(\ref{ricci_Min_a}). Doing this we get
\begin{eqnarray} \label{warp_eq_1}
\widetilde{\triangledown}^2(4A+B)-3\widetilde{g}^{mn}\partial_nB\partial_m(4A+B)
&=&
e^{-2A-2B}\frac{G_{mnp}\bar{G}^{mnp}}{12\textrm{Im}\widetilde{\tau}}+e^{-10A-4B}\partial_m\alpha\partial^m\alpha
\nonumber\\
&&+\frac{k^2_{10}}{2}e^{-2A-2B}(T^m_m-T^{\mu}_{_\mu})^{loc}
\end{eqnarray}
 On the other hand  using
(\ref{Ricci_Min}) in (\ref{ricci_Min_a}), one gets
\bg \label{BHfactorA1}
R_t^t-R_x^x=0
\nd
which in turn would immediately imply
\bg\label{BHfactorA}
\widetilde{\triangledown}^2 B-3\widetilde{g}^{mn} \partial_m B \partial_n B=0
\nd
Now let's look at the Bianchi identity. Using (\ref{dF5}) in
(\ref{bianchi}) gives
\begin{eqnarray}
\tilde{\triangledown}^2\alpha -3e^{-2A-2B}\partial_m
B\partial^m\alpha&=&e^{2A-B}\frac{\ast_6 \bar{G}_3\cdot
G_3}{12i\textrm{Im}
\widetilde{\tau}}+2e^{-6A-3B}\partial_m e^{4A+B}\partial^m\alpha.
\end{eqnarray}
Subtracting it from (\ref{warp_eq_1}) one gets the following
\begin{eqnarray}\label{GKP_BH}
\widetilde{\triangledown}^2(e^{4A+B}-\alpha)&=&\frac{e^{2A-B}}{24\textrm{Im}\widetilde{\tau}}|\textrm{i}
G_3-\ast_6G_3|^2+e^{-6A-3B}|\partial (e^{4A+B}-\alpha)|^2\nonumber\\
&&+3e^{-2A-2B}\partial_mB\partial^m(e^{4A+B}-\alpha)+\textrm{local
source}
\end{eqnarray}

Next we consider the internal components of Einstein equations.
Ricci tensor for the $x^m, m=4,..,9$ directions take the following form
\bg \label{Rmn_fangA}
R_{mn}&=&\widetilde{R}_{mn} + \widetilde{g}_{mn} \widetilde{\triangledown}^2\left(A + B\right)-3\widetilde{g}_{mn}\widetilde{g}^{\lambda k}\partial_\lambda B
\partial_k
\left(A + B\right) \nonumber\\
&+& 3\widetilde{\triangledown}_m\partial_n B +
\partial_mB\partial_nB -8\partial_mA\partial_nA -
2\partial_{(m}A\partial_{n)}B \nd where
$\widetilde{\triangledown}_m$ is the covariant derivative given by
\bg \widetilde{\triangledown}_mV_c=\partial_m
V_c-\widetilde{\Gamma}_{mc}^b V_b \nd for any vector $V_b$. Here
$\widetilde{R}_{mn}$ is the Ricci tensor and
$\widetilde{\Gamma}_{mc}^b$ is the Christoffel symbol for the metric
$\widetilde{g}_{mn}$.
Using (\ref{warp_eq_1}) and (\ref{BHfactorA}) we find
\begin{eqnarray}\label{tricci}
\tilde{R}_{mn}&=&
 -g_{mn}\frac{G_3\cdot\bar{G}_3}{24\textrm{Im}
\widetilde{\tau}}+\frac{G_{mab}\cdot\bar{G}_n^{ab}}{4\textrm{Im}\widetilde{\tau}}+\frac{\partial_m\widetilde{\tau}
\partial_n\bar{\widetilde{\tau}}}{2\mid\textrm{Im}\widetilde{\tau}\mid^2}\nonumber\\
&& -\frac{1}{2}e^{-8A-2B}\partial_m\alpha\partial_n\alpha+8\partial_m A\partial _n A\nonumber\\
&&-3\widetilde{\triangledown}_m\partial_n B - \partial_mB\partial_nB
 + 2\partial_{(m}A\partial_{n)}B
\end{eqnarray}

To solve for these equations we must first find the three-form flux.
For simplicity we will ignore the running of the $\widetilde{\tau}$ field, that
is  $d\widetilde{\tau}=0$. We further take $C_0=0$ and $e^{\phi}=g_s$.  With
constant $\widetilde{\tau}$ and considering fluxes on the deformed cone, $G_3=F_3-\widetilde{\tau} H_3$ must be closed  everywhere since we are
ignoring all localized sources i.e. $S_{\rm loc}=0$ in (\ref{Action}). Then the three form flux equation becomes
\begin{equation}\label{G3eom}
d(e^{4A+B}\ast_6G_3-i\alpha G_3)=0
\end{equation}
Written in terms of $F_3$ and $H_3$, the above equation  becomes
\begin{eqnarray}\label{FHeom}
dC_1=0, ~~C_1\equiv e^{4A+B}\ast_6F_3-\alpha e^{-\phi}H_3\nonumber\\
dC_2=0, ~~ C_2\equiv e^{4A+B}\ast_6e^{-\phi}H_3+\alpha F_3
\end{eqnarray}
where $C_1$ and $C_2$ are related by
\begin{eqnarray}\label{C1C2}
\ast_6C_2=\left(\alpha-\frac{e^{8A+2B}}{\alpha}\right)\ast_6F_3+\frac{e^{4A+B}}{\alpha}C_1.
\end{eqnarray}

Our goal is to solve the system of all Einstein equation and field equations exactly. To do this
we take the simplest case where we assume all the unknown functions $A$, $B$, $a$, $b$ and $\alpha$
 only depends on radial direction $r$. In the non-AdS extremal limit and the AdS non-extremal limit, all these functions are known, so we can use the known  forms as a starting point.
 We use the following ansatz for the warp factors
 \bg\label{ans-AB}
 e^{-4A}\equiv h(r)&=&\frac{N\alpha'^2}{r^4}\left(1+ \frac{81g_sM^2}{8 N} {\rm log} \Big(\frac{r}{r_l}\Big)+ A_1(r)\right)\nonumber\\
 e^{2B}\equiv g(r)&=& 1-\frac{\tilde{r_h}^4}{r^4}+G(r)
 \nd
where $r_l$ is some scale determining boundary condition of $h(r)$ and $\tilde{r}_h$ is the Schwarzschild horizon i.e. 
the black hole horizon in the limit $M=0$. We will later discuss the relation between
the actual horizon $r_h$ and $\tilde{r}_h$. But for now note that
 for  $F_3$ to recover the Klebanov-Tseytlin solution in the limit $B=0$, the
 corrections must be proportional to the black hole horizon $r_h$. On
 the other hand, these corrections must vanish as $r\mapsto \infty$.
 However $F_3$ must not depend on $r$ since D5 brane is orthogonal
 to the radial direction.
 Thus we argue that  $F_3$ remains the same
 \begin{eqnarray}\label{F3}
 F_3=\frac{M\alpha'}{2}g^5\wedge \omega_2
 \end{eqnarray}
 where $\omega_2=g^1\wedge g^2+g^3\wedge g^4$. On the other hand, using this form in
 (\ref{FHeom}) gives
\begin{eqnarray} \label{H3full}
\alpha H_3&=&\Big(\frac{3Me^{\phi}\alpha'e^{4A}}{2r} \sqrt{\frac{a}{b}}dr\wedge \omega_2 -e^{\phi}C_1\Big)
\end{eqnarray}
where we have used the ansatz (\ref{g'}),(\ref{ans-AB})  and ignored terms of order
${\cal O}\left(r_hg_sM^3/N^2\right)$, which is a good approximation in the limit $N\gg M$. If we assume $C_1=c_1(r)dr\wedge \omega_2$, this automatically gives $dH_3=0$. Now from the closure of $C_2$ and using (\ref{C1C2}), we find
\begin{eqnarray}
c_1(r)=\frac{3M\alpha'}{2r}\sqrt{\frac{a}{b}}\Big(e^{4A}-\frac{\alpha
(\alpha-\alpha(r_h))}{e^{4A+2B}}\Big)
\end{eqnarray}
so we get
\begin{eqnarray} \label{H3}
e^{-\phi}H_3=\frac{3M\alpha'
e^{-4A-2B} (\alpha-\alpha(r_h))}{2r}\sqrt{\frac{a}{b}}dr\wedge
\omega_2
\end{eqnarray}
We further compute $G_3\cdot \bar{G}_3$
\begin{eqnarray}
G_3\cdot \bar{G}_3dx^6&=&\frac{3! G_3\wedge \ast_6
\bar{G}_3}{\sqrt{g_6}}\nonumber\\
&=&\frac{18M^2\alpha'^2e^{-B}\left(1+ {\cal O}(r_hg_sM^2/N)\right)}{4r}\sqrt{\frac{a}{b}}(1+(\alpha-\alpha(r_h))^2e^{-8A-2B})\sin\theta_1\sin\theta_2/\sqrt{g_6}dx^6\nonumber\\
\sqrt{g_6}&=&\frac{e^{-6A-B}r^5\sqrt{ab}b^2\left(1+ {\cal O}(r_hg_sM^2/N)\right)\sin\theta_1\sin\theta_2}{108}
\end{eqnarray}
so
\begin{eqnarray} \label{G3^2}
G_3\cdot
\bar{G}_3=\frac{486M^2\alpha'^2e^{6A}}{r^6b^3}(1+(\alpha-\alpha(r_h))^2e^{-8A-2B})
\end{eqnarray}
where we have ignored terms of ${\cal O}\left(r_hg_sM^3/N^2\right)$ that arise due to the deformation of the base $\tilde{g'}_{pq}$.

On the other hand, $G_{mab}\bar{G}_n^{ab}$ is
\begin{eqnarray}
G_{mab}\bar{G}_n^{ab}&=&F_{mab}F_n^{ab}+e^{-2\phi}H_{mab}H_n^{ab}\nonumber\\
&=&\frac{g_{mn}}{3}(\ast F_3)\cdot(\ast F_3)-(\ast F)_{mab}(\ast F)_n^{ab}+e^{-2\phi}H_{mab}H_n^{ab}\nonumber\\
&=&g_{mn}\frac{27M^2\alpha'^2e^{6A}}{r^6b^3mf}-(\ast F)_{mab}(\ast
F)_n^{ab}+e^{-2\phi}H_{mab}H_n^{ab}
\end{eqnarray}
While $T_{mn}\equiv -(\ast F)_{mab}(\ast
F)_n^{ab} +e^{-2\phi}H_{mab}H_n^{ab}$ only has following non-zero
components
\begin{eqnarray}
T_{rr}&=&\frac{9M^2\alpha'^2e^{-2B}}{4r^2}\frac{a}{b}\Big(e^{-8A-2B}(\alpha-\alpha(r_h))^2-1\Big)\frac{72}{e^{-4A}r^4b^2}\nonumber\\
&=&g_{rr}\frac{162M^2\alpha'^2e^{6A}}{r^6b^3}\Big(e^{-8A-2B}(\alpha-\alpha(r_h))^2-1\Big)\nonumber\\
T_{\phi_i\phi_i}&=&\frac{9M^2\alpha'^2e^{-2B}}{4r^2}\frac{a}{b}\Big(e^{-8A-2B}(\alpha-\alpha(r_h))^2-1\Big)
\frac{6\sin^2\theta_i}{e^{-4A-2B}r^2ab}\nonumber\\
&=&g_{\phi_i\phi_i}\frac{243M^2\alpha'^2e^{6A}\sin^2\theta_i}{r^6b^3(2\cos^2\theta_i^2+3\sin^2\theta_i)}\Big(e^{-8A-2B}(\alpha-\alpha(r_h))^2-1\Big)\nonumber\\
 T_{\theta_i\theta_i}&=&\frac{9M^2\alpha'^2e^{-2B}}{4r^2}\frac{a}{b}\Big(e^{-8A-2B}(\alpha-\alpha(r_h))^2-1\Big)\frac{6}{e^{-4A-2B}r^2ab}\nonumber\\
&=&g_{\theta_i\theta_i}\frac{81M^2\alpha'^2e^{6A}}{r^6b^3}\Big(e^{-8A-2B}(\alpha-\alpha(r_h))^2-1\Big)
\end{eqnarray}
where again we have ignored terms of ${\cal O}\left(r_hg_sM^3/N^2\right)$ that appear due to the deformation of metric of base
$\tilde{g'}_{pq}$.

Using the above forms of the fluxes, the only {\it non-trivial} internal Einstein equations\footnote{By trivial Einstein equations we mean
Ricci tensor, $\tilde{R}_{mn}=0$ equations.} (\ref{tricci}) are
\begin{eqnarray} \label{Ricci_int}
\tilde{R}_{rr}&=&g_{rr}\frac{297Q}{8}-\frac{1}{2}e^{-8A-2B}\partial_r\alpha\partial_r\alpha+8\partial_r A\partial _r A\nonumber\\
&&-3\widetilde{\triangledown}_r\partial_r B - \partial_rB\partial_rB
 + 4\partial_rA\partial_rB\nonumber\\
\tilde{R}_{\psi j}&=&-g_{\psi j}\frac{27Q}{8}+3\tilde{\Gamma}^r_{\psi j}\partial_rB\nonumber\\
\tilde{R}_{\phi_1\phi_2}&=&-g_{\phi_1\phi_2}\frac{27Q}{8}+3\tilde{\Gamma}^r_{\phi_1\phi_2}\partial_rB\nonumber\\
\tilde{R}_{\phi_i\phi_i}&=&g_{\phi_i\phi_i}\frac{27(15\sin^2\theta_i-2\cos^2\theta_i)Q}{8(2\cos^2\theta_i+3\sin^2\theta_i)}+3\tilde{\Gamma}^r_{\phi_i\phi_i}\partial_rB\nonumber\\
\tilde{R}_{\theta_i\theta_i}&=&g_{\theta_i\theta_i}\frac{135Q}{8}+3\tilde{\Gamma}^r_{\theta_i\theta_i}\partial_rB\nonumber\\
\end{eqnarray}
where
$Q=M^2\alpha'^2e^{6A}(e^{-8A-2B}(\alpha-\alpha(r_h))^2-1)/(r^6b^3\textrm{Im}
\tau)$ and $j=\psi$, $\phi_i$.

We will now solve the set of Einstein equations (\ref{warp_eq_1}),(\ref{BHfactorA}) and (\ref{Ricci_int}) along with the trivial Einstein
equations and the  Bianchi identity
(\ref{GKP_BH}). Note that with the ansatz (\ref{inmated}) for the metric, the three form fluxes $H_3, F_3$ which satisfy the flux
equation (\ref{G3eom}) are already determined and given by (\ref{F3}) and (\ref{H3}). Thus, once equations
(\ref{warp_eq_1}),(\ref{BHfactorA}), (\ref{GKP_BH}) and (\ref{Ricci_int}) are solved, we have a complete set of solution as long as the
trivial Einstein equations are also solved. To proceed
with the analysis, we make the following ansatz for the five form flux strength
\bg \label{alphaa}
\alpha= t(r) e^{4A},~~~~~~ t(r)\equiv 1+ t_1(r)= 1+ {\cal O}(r_hg_sM^2/N)
\nd
which is again consistent with the extremal limit (i.e. $r_h=0$) and the AdS limit $M=0$. The ${\cal O}(r_hg_sM^2/N)$ term is exactly
consistent with the ansatz for the warp factors (\ref{ans-AB}) and the internal metric (\ref{inmated}), (\ref{g'}).

Also observe that the
above ansatz (\ref{alphaa}) along with the warp factors (\ref{ans-AB}) reduce to the well known Klebanov-Witten black hole solutions in the
limit $M=0$. In fact, for $M=0$, the horizon is given by $\tilde{r}_h$ with $A_1(r)=G(r)=0$ in(\ref{ans-AB})
and we have a Schwarzschild black hole in
$AdS_5\times S^5$. We will refer to $\tilde{r_h}$ as the Schwarzschild horizon while $r_h$
denotes the black hole horizon in the presence of three form flux and other sources.
With no black hole but $M\neq 0$, our ansatz reduces to the extremal limit of Klebanov-Tseytlin model with $\tilde{r_h}=r_h=B=A_1=G=0$. With
these limiting behavior and the scaling of the warp factors, it is reasonable that the functions $A_1(r), G(r)$ appearing in
(\ref{ans-AB}) should have the following scaling
\bg \label{A1G}
A_1&=& {\cal O}(r_h g_sM^2/N)\nonumber\\
G&=&{\cal O}(r_h g_sM^2/N)
\nd

With the above ansatz for the warp factors and $\alpha$, the system of equations can be further simplified. Using the expansions
(\ref{ans-AB}), (\ref{alphaa}) and
(\ref{A1G}), we find for $r\neq \tilde{r}_h$ , expanding up to linear order in $g_sM^2/N$ \footnote{
The Taylor series in  $g_sM^2/N$ is well defined $\forall r$ except $r=
\tilde{r}_h$ and thus the expansion is valid away from but arbitrarily close to $r=\tilde{r}_h$.}
\bg
e^{-8A-2B}(\alpha-\alpha(r_h))^2=1-\frac{\tilde{r}_h^4}{r^4}+{\cal O}(r_h g_sM^2/N)
\nd
Now  the Taylor series in $\tilde{r}_h/r$ away from $r=\tilde{r}_h$\footnote{The Taylor series for $\tilde{r}_h/r$  is well defined even at
$r=\tilde{r}_h$, for $M\neq 0$.}, keeping only up to linear order terms in ${\cal O}(g_sM^2/N)$
 gives 

\bg
g_{pq} Q&\sim& \frac{\tilde{r}_h^{4}}{r^{4}} {\cal O}(g_sM^2/N) \nonumber\\
g_{rr} Q &\sim& \sum_{k=1}\frac{\tilde{r}_h^{4k}}{r^{4k+2}} {\cal O}(g_sM^2/N)
\nd
  But in the limit $N\gg M$, we have ${\cal O}(g_sM^2/N) \ll 1$ and for
$\tilde{r}_h/r<1$, we can ignore $g_{mn} Q$ all together. This
drastically reduces  equations in (\ref{Ricci_int}) to

\begin{eqnarray} \label{Ricci_int_simple}
\tilde{R}_{rr}&=&-\frac{1}{2}e^{-8A-2B}\partial_r\alpha\partial_r\alpha+8\partial_r A\partial _r A\nonumber\\
&&-3\widetilde{\triangledown}_r\partial_r B - \partial_rB\partial_rB
 + 4\partial_rA\partial_rB\nonumber\\
\tilde{R}_{ij}&=&3\tilde{\Gamma}^r_{i j}\partial_rB\nonumber\\
\end{eqnarray}
In fact, the above equation simplifies further, with the following observation: in the limit $M=0$, (\ref{Ricci_int_simple}) is trivially
satisfied for our ansatz of the warp factor (\ref{ans-AB}) along with the scalings (\ref{A1G}),(\ref{alphaa}) and (
\ref{g'}). For $M\neq 0$, the right side of the
equation has terms of order
${\cal O}(\tilde{r}_h^{4k}/r^{4k}){\cal O}(g_sM^2/N) $, if we ignore the deformation of the base (that is $a=b=1, G=0$ and $\tilde{g'}_{pq}=0$)
and consider $t_1=A_1=0$. On the other hand, the left side $\tilde{R}_{mn}$ can have terms of 
${\cal O}(\tilde{r}_h^{4+l}/r^{4+l}){\cal O}(g_sM^2/N) $ if the
base of the cone along with $t_1, A_1, G$ have terms of same order. This means, to solve (\ref{Ricci_int}) everywhere from horizon to boundary
along with the {\it trivial} internal Einstein equations,
we expect the following scalings :

\bg \label{gpq}
{\cal F}_1\sim  \sum_{l=0}^{\infty}\left(\frac{\tilde{r}_h}{r}\right)^{4+l} {\cal O}\left(\frac{g_sM^2}{N}\right)\nonumber\\
\tilde{g'}_{pq}\sim \sum_{l=0}^{\infty}\left(\frac{\tilde{r}_h}{r}\right)^{4+l} {\cal O}\left(\frac{g_sM^2}{N}\right)
\nd
where ${\cal F}_1\equiv A_1(r), t_1(r), a_1(r), b_1(r)$ or $G(r)$.
Away from the Schwarzschild horizon i.e $r>\tilde{r}_h$ such that ${\cal O}(g_sM^2/N)\sim {\cal O}(\tilde{r}_h^4/r^4)< 1$, we can ignore
${\cal O}(g_sM^2/N){\cal O}(\tilde{r}_h^4/r^4)$ and then (\ref{Ricci_int}) is trivially solved. In fact, ignoring
$A_1(r), t_1(r), a_1(r), b_1(r)$ and $G(r)$ away from the horizon gives that  all the equations (\ref{warp_eq_1}),(\ref{BHfactorA}),(\ref{Ricci_int}) along with the Bianchi identity
(\ref{GKP_BH}) are satisfied with the following solution
\bg \label{warpBD}
e^{-4A}&=&h_0(r)=\frac{N\alpha'^2}{r^4}\left(1+ \frac{81g_sM^2}{8 N} {\rm log} \frac{r}{r_l}\right)\nonumber\\
 e^{2B}&\equiv& g_0(r)= 1-\frac{\tilde{r_h}^4}{r^4}
 \nd

 Now what is the relation between the real horizon $r_h$ and the Schwarzschild horizon $\tilde{r}_h$? To answer this,
note that
since $N\gg M$, we do not expect $G$ to be very large even at the horizon. This is also consistent with the scaling in (\ref{gpq}).
 This means, the true horizon $r_h$ is close to the Schwarzschild
horizon $\tilde{r}_h$. More precisely we expect\footnote{We expect $D5$ branes to increase the mass of the black hole and thus 
$r_h>\tilde{r}_h$. This is also consistent with our numerical analysis in \cite{fangmia}.}

\bg
r_h= \tilde{r}_h\left(1+ {\cal O}(g_sM^2/N)\right)
\nd

Thus $r>\tilde{r}_h$ also means $r>r_h$ and hence the warp factors (\ref{warpBD}) with ignoring ${\cal F}_1, \tilde{g'}_{pq}$ 
works exactly away from the horizon. At the horizon, the Ricci tensor equations
(\ref{Ricci_int_simple}) are not valid and we have to incorporate the deformation of the base metric $\tilde{g'}_{pq}$ in solving the
system of equations. In fact exactly at the horizon we can see that $g_{mn}Q$ is no longer ignorable up to linear order since
$\tilde{r}_h/r\sim 1$. If $\tilde{g'}_{pq}=0$ and we consider the near horizon region, there are more than two independent equations in
(\ref{Ricci_int})- resulting in  more equations than the number
of independent functions i.e $A_1,G,t_1,a_1,b_1$.  Thus we must consider the deformations $\tilde{g'}_{pq}$ near the horizon to solve all
the Einstein equations in (\ref{Ricci_int}) and the flux equations exactly.
The form of
the ansatz (\ref{gpq}) is precisely chosen to solve all the equations in (\ref{Ricci_int}) along with the trivial equations.

As long as we are away from the horizon, the details of the metric $\tilde{g'}_{pq}$ and functions $A_1(r), t_1(r), a_1(r), b_1(r),G(r)$
will not effect  our calculations since all these functions are second or higher order in $\gamma \sim {\cal O}(g_sM^2/N)\sim {\cal
O}(r_h^4/r^4) $. Note that the flux equation (\ref{G3eom}) is solved exactly in the entire region $r_h \le r\le \infty$
up to linear order in ${\cal O}(r_hg_sM^2/N)$ and this equation does not involve  $\tilde{g'}_{pq}$. On the other hand,
using (\ref{alphaa}) in (\ref{H3full}), one would obtain that $ H_3$
will get contribution from $\tilde{g'}_{pq}$. But what enters into the system of equations are not just $H_3$ but the tensor $T_{mn}$ and
$g_{mn} G_3\cdot \bar{G}_3$, which up to linear order and away from the horizon give rise to the simplified equation (\ref{Ricci_int_simple})-
which does not involve $\tilde{g'}_{pq}$.

In summary, we have been able to obtain the exact form of  flux and metric in the presence of a black hole. The three form  flux is given by
(\ref{F3}) and (\ref{H3}), while the scalar function that gives rise to the five form flux is given in (\ref{alphaa}). The modification $t_1$
 to
$\alpha$ from the extremal limit, the modification of the warp factor $A_1, G$ and  deviation of the
internal metric is described by functions $a_1,b_1$ and $\tilde{g}'_{pq}$- all of which take the form (\ref{gpq}). Thus if $r_h>\tilde{r}_h$,
we have exact expressions for the metric and the fluxes in the entire region $r_h \le r\le \infty$, up to linear order in ${\cal O}(g_sM^2/N)$. Most importantly, using
 these solutions and the solution for the warp factor (\ref{ans-AB}), one can exactly compute the three form flux strength
 up to linear order in ${\cal O}(g_sM^2/N)$- a result that allows one to compute the on shell action which will be shown in section \ref{NCFT2}.

\section{Phase Transitions}\label{PT}
Phase transitions of four dimensional $SU(N)$ gauge theory at finite temperature can be realized
by spontaneous breaking of the center symmetry
$Z_N$ along the Euclidian time circle. The order parameter associated with this
 symmetry is the temporal Polyakov loop operator
 \begin{eqnarray}
 W_0=\frac{1}{N}\textrm{Tr}\;P\;{\rm exp}\left(i\int_0^{\beta} d\tau \lambda_a A_0^a \right)
 \end{eqnarray}
 where $\beta$ is the periodicity of Euclidian time, $P$ denotes time ordering and $\lambda_a$ are the Gell-Mann matrices for $SU(N)$.
 In the confined phase, $Z_N$ symmetry is preserved
and $\langle W_0\rangle $ is zero. In the deconfined phase,
$Z_N$ symmetry is spontaneously broken and $\langle W_0\rangle \neq0$. 

In \cite{Mia:2010tc}, we computed $\langle W_0\rangle$ in both extremal and
non-extremal geometries following the procedure of \cite{Mal-2}-\cite{reyyee2}. We found that for the extremal geometry, $\langle W_0\rangle=0$ while for the black hole geometry, $\langle
W_0\rangle\neq 0$. Although the calculation was done in UV complete geometry, it is straight forward to demonstrate that for a
Klebanov-Strassler geometry with a deformed cone and constant warp factor $h(\rho)\sim constant\neq 0$ near $\rho\sim 0$, the Nambu-Goto
action $S_{NG}\sim \beta F_{Q\bar{Q}}\sim \beta {\cal D}$, i.e. it  linearly increases with inter quark separation ${\cal D}$, for large ${\cal
D}$. Since 
$\langle W_0 W_0^\dagger \rangle \sim {\rm exp}(-\beta F_{Q\bar{Q}})$ and the free energy of a single quark is obtained in the limit
$F_Q=\lim_{{\cal D}\rightarrow \infty} F_{Q\bar{Q}}/2$, we get $\langle W_0\rangle\sim {\rm exp}(-\beta F_{Q})=0$. For the non-extremal geometry, even with non-UV complete scenarios, one can follow the procedure in \cite{Mia:2010tc} and obtain 
that $\langle
W_0\rangle\neq 0$. 

Thus the extremal geometry $X^1$ corresponds to zero order parameter while black hole geometry $X^2$
 describes non-zero order parameter. Furthermore, as already discussed in section \ref{extm},  using (\ref{pE}) 
 one directly obtains that $X^1$ corresponds to zero entropy while $X^2$ has black holes and non-zero entropy. Thus $X^1$ is the gravity dual to the
 confined phase of the gauge theory  while $X^2$ describes the deconfined phase. Note that although $X^1$ corresponds to zero entropy, it can
 describe the gauge theory at non-zero temperature and we can obtain the free energy of the dual gauge theory by using (\ref{FreeE}) with the
  on shell action for $X^1$. This is consistent with the understanding that confined phase can exist not only at zero temperature, but up to
  some non-zero critical temperature. However, at a given temperature, only one description is favored.  By comparing the on shell actions for $X^1$ and $X^2$
  we can obtain which geometrical description is preferred and  study the Hawking-Page like phase
transition between these two. But before we do that, we first study the holographic descriptions of conformal theories and demonstrate how
there is no  phase transitions.

\subsection{Conformal field theory}
For the $\mathcal{N}=4$ Yang-Mills theory on 4D Minkowski space, there are two gravity solutions too, one is 
$AdS_5\times S^5$ ($X^1$) and the other is
black hole $AdS_5\times S^5$ ($X^2$), with the Euclidean metric in Einstein frame is
\begin{eqnarray}\label{AdSmetric}
ds^2&=&\frac{r^2}{{\cal L}^2}\Big[\left(1-\frac{\tilde{r}_h^4}{r^4}\right)d\tau^2+\vec{dx}^2\Big]+
\frac{{\cal L}^2}{r^2\left(1-\frac{\tilde{r}_h^4}{r^4}\right)} dr^2+{\cal L}^2d\Omega^2_5
\end{eqnarray}
where $\tilde{r}_h$ is the black hole horizon and ${\cal L}^4=4\pi N_c \alpha'^2$ where $N_c$ is the number of $D3$ branes in the dual gauge
theory\footnote{Using (\ref{Neff}) with the surface integral over $S^5$ and $\tilde{F}_5=(1+\ast) dh^1\wedge dt\wedge dx\wedge dy \wedge dz$,
$h=\frac{{\cal L}^4}{r^4}$, one readily gets $N_c=\frac{{\cal L}^4}{4\pi \alpha'^2}$}. 
Note that if
$R, R_s$ are the Ricci scalar in Einstein and  string frame, we have $R_s=R/\sqrt{g_s}$ with $e^\phi=g_s$ being the string coupling. Denoting
string frame metric as $g_{s MN}$ and Einstein metric as $g_{MN}$, we have  
Ricci scalar for the AdS direction as $g^{ab}R_{ab}\sim -\frac{1}{{\cal L}^2}, a,b=0,..,4$ and in the string frame 
$g^{ab}_s R_{s ab}\sim -\frac{1}{ L^2}, L\equiv g_s^{1/4} {\cal L}$ . For supergravity to be a good approximation we need Ricci scalar to be small in
string frame, which means we need $L^2=\sqrt{g_sN_c}\alpha'\gg 1$. On the other hand we want to ignore string interactions, which means
$g_s\rightarrow 0$. Thus the limit $\sqrt{g_sN_c}\alpha'\gg 1$ can be obtained with $N_c\rightarrow \infty$ and classical gravity with the metric
(\ref{AdSmetric}) will describe
the dual geometry. Observe that when $\tilde{r}_h=0$, we recover the $AdS_5\times S^5$ metric without any black holes.

 The {\it bulk} on shell supergravity action for both $X^1$ and $X^2$ is the same
\begin{eqnarray}
I_{\rm SUGRA}=\frac{1}{2\kappa_{10}^2}\int d^{10}x \sqrt{G}\Big(R-\frac{|F_5|^2}{4\cdot5!}\Big)=0
\end{eqnarray}
So the action difference should come from the Gibbons-Hawking (GH) term \cite{Gibbons-Hawking,Henningson:1998gx}
\begin{eqnarray}
I_{\rm GH}=\frac{1}{\kappa_{10}^2}\int_{M_9}d^9x\sqrt{\textrm{det}{\tilde{g}}}(D_{\mu}n^{\mu})
\end{eqnarray}
where $M_9$ is the boundary at $r={\cal R}$, $\tilde{g}$ is the metric induced on $M_9$, $n^{\mu}$ is a unit normal vector
to $M_9$ and $D_{\mu}$ is the covariant derivative.

It is straight forward to compute the GH term for $AdS_5\times S^5$ with and without black holes. We find the GH term is
\begin{eqnarray}
I_{\rm GH}^2=\frac{\beta_2V_3 \pi^3}{\kappa_{10}^2}\lim_{\cal R\rightarrow \infty}\left(4{\cal R}^4-2r_h^4\right)
\end{eqnarray}
for black hole $AdS_5\times S^5$,  and
\begin{eqnarray}
I_{\rm GH}^1=\frac{4\beta_1V_3 \pi^3}{\kappa_{10}^2}\lim_{\cal R\rightarrow \infty} {\cal R}^4
\end{eqnarray}
 for $AdS_5\times S^5$ and we have used that volume of $S^5$ is $\int d\Omega_5= \pi^3$. 
 
 However, the extremal geometry $X^1$  only describes zero temperature. The temperature described by the geometry is obtained from 
  the periodicity of Euclidean time and for $X^1$ which is Poincar\'e AdS, this period cannot be arbitrary. This is because
  Poincar\'e AdS metric 
 (obtained by setting $\tilde{r}_h=0$ in (\ref{AdSmetric}))   
 is singular at $r=0$. Near $r=0$, with $V_1=r^2/{\cal L}^2$ the $(r,\tau)$ part of the metric takes the following form 
 \bg
 ds^2=\frac{1}{V_1 V_1'^2}\left(dV_1^2+ V_1^2 V_1'^2 d\tau^2\right)
 \nd
 where $V_1'=\frac{dV}{dr}$. The singularity at $V_1=0$ is really  the singularity of the origin of polar coordinates and the metric is smooth
 and complete, 
 provided $|V_1'|d\tau\equiv
 d\theta$, where $\theta$ is the polar angle. This readily gives that period  $\beta_1=\int d\tau= 1/|V_1'| \int d\theta\rightarrow \infty $ as 
 $V_1\rightarrow 0$. Thus
 the corresponding temperature is 
 
\bg T_1=\beta_1^{-1}=0 
\nd
On the other hand, 
 the Euclidean black hole metric is singular at $r=\tilde{r}_h$. The singularity can be realized as that of polar coordinates 
  with the change of variable $V_2=\frac{r^2}{L^2}\left(1-\frac{\tilde{r}_h^4}{r^4}\right)$ and looking at the metric near the sigular point 
  $V_2=0$. This way the metric is smooth and complete if and only if 
\bg \label{beta2}
\beta_2=\frac{\pi {\cal L}^2}{\tilde{r}_h}
\nd
which depends on the horizon $\tilde{r}_h$. 
Also the temperature of the field theory at
hypersurface $\rho={\cal R}$ of geometry $X^2$ is now redshifted due
to the presence of the black hole \footnote{The $g_{tt}$ and
$g_{cc},c=x,y,z$ component of the Minkowski metric at hypersurface
$r={\cal R}$ are now distinct. The difference is dependent on
${\cal R}$ and thus the local temperature at $r={\cal R}$ is
dependent on ${\cal R}$}
\bg \label{T}
T_2({\cal R})=\frac{\beta_2^{-1}}{\sqrt{\left(1-\frac{\tilde{r}_h^4}{{\cal R}^4}\right)}}
\nd

In summary, the Poincar\'e AdS metric without a black hole only describes the vacuum phase and corresponds to zero temperature. On the other
hand, for any non-zero temperature, ${\cal N}=4$ SUSY gauge theory is described by the black hole geometry. Since there is no description in
terms of distinct geometries for a given non-zero temperature, there is no Hawking-Page transition.    

Now we would like to find the thermodynamic parameters for the $SU(N_c)$ conformal field theory.
But since the black hole action $I^2_{\rm GH}$ is divergent, we need to regularize it first. The divergent part $\sim \beta_2 {\cal R}^4$ is
dependent on temperature and  thus the counter term will also be dependent on it. Also observe that the regularization scheme adopted in 
\cite{Hawking:1982dh,Witten:1998zw} identifies $\Delta S$ as the regularized gravity action. Hence regularization is not just subtraction of
 the infinite part $\sim \beta_2 {\cal R}^4$, but also a finite part arising from the
vacuum action. There is one crucial difference between the case studied in \cite{Hawking:1982dh,Witten:1998zw} and 
the Poincar\'e AdS geometry. 
Essentially, Hawking-Page considered global AdS geometry, where  without a black hole, the  geometry is regular and 
the periodicity of Euclidean
time can be arbitrary. There, at a given temperature there are two descriptions, black hole and no black hole
geometries  and  $\Delta S$ can describe the
excess of energy due to presence of the black hole. For Poincar\'e AdS geometry, the action $S^1$ only describes zero temperature, so
$\Delta S$ is only meaningfull at zero temperature, where it is trivial. At any non-zero temperature, $\Delta S$ {\it does not}
 describe the excess energy and thus cannot be identified with the renormalized action for the black hole.
 
 However, the
dual conformal thermal gauge theory has free energy that scales as $\sim T^4$ and entropy that scales as $T^3$. To extract this information
from our Poincar\'e AdS black hole geometry,  special care is needed while regulatizing the action.
 In particular, one needs to holographically renormalize
the action  following the procedure of \cite{Henningson:1998gx} and write the regularized action as a functional of the induced metric at the
boundary to find the stress tensor of the dual gauge theory. 

There is an alternative approach to obtain the thermodynamic state functions. 
Black hole geometries have entropy proportional to one quarter the surface area of the horizon \cite{Bekenstein:1973ur, Hawking:1974sw}. 
We can directly evaluate the black hole entropy for the ten dimensional $AdS^5\times S^5$ black hole geometry  using Wald's formula 
\cite{wald1}-\cite{wald4}
where the gravity Lagrangian is  ${\cal L}_{\rm bulk}=\frac{1}{2\kappa_{10}^2}\left(R-\frac{|F_5|^2}{4\cdot5!}\right)$.
The result is 
\begin{eqnarray}
s=\frac{2\pi^4 \tilde{r}_h^3 {\cal L}^2V_3}{\kappa_{10}^2}=\frac{1}{2} N_c^2 \pi^2 V_3 T^3
\end{eqnarray}    
where $T\equiv T_2({\cal R}=\infty)$ and we have used definition of $\kappa_{10}^2=2^6\pi^7 \alpha'^4$.
Knowing the entropy, we can of course obtain the free energy and find the appropriate counter terms that would lead to this free energy. The
result is 
 \bg
 F&=&=-\frac{1}{8}V_3\pi^2 N_c^2 T^4=T(I^2_{\rm SUGRA}+I^2_{\rm GH}+ I^2_{\rm counter})\nonumber\\
 I^2_{\rm counter}&=&-\frac{4{\cal R}^4-3/2 \tilde{r}_h^4}{\kappa_{10}^2}\beta_2 V_3 \pi^3
 \nd  

The result for entropy and free energy is consistent with the finite temperature conformal field theory since they have the correct scaling
with temperature. Also note that the entropy  $s=3/4 s_0$ where
$s_0$ is the entropy for weakly coupled ${\cal N}=4$ SYM theory.  We
  will not discuss this discrepancy here, please refer to \cite{Gubser:1996de}, \cite{Gubser:1998nz} and \cite{Fotopoulos:1998es}.
  
Observe that the discussion about the phase transition about the conformal field theory is different from the case discussed in 
\cite{Witten:1998zw}. 
There the field theory lives on $S^1\times S^3$, with circumference $\beta$ and $\beta'$ respectively. Thus the field theory is no longer 
conformally invariant and depends on the scale $\beta/\beta'$. In the large $N$ limit there can be a phase transition as a function of $\beta/\beta'$.

\subsection{Non-conformal field theory}\label{NCFT}

Now we turn to the phase transition of non-conformal field theory.
We have discussed dual gravity solutions of the non-conformal field theories in section $2$.
 The zero-temperature solution which is discussed in section
\ref{extm} is UV completed. The geometry is a warped deformed
conifold in the IR and asymptotically AdS in the UV which ensures that the theory is holographically renormalizable.
When energy flows from UV to IR as
$r$ gets smaller and smaller, the effective number of D3 brane is
cascaded away, leaving only D5 branes in the IR and the theory is
confined. In the previous section we discussed the
finite-temperature solution but we didn't bother to UV complete it.
However notice that the same AdS cap can be added to the black hole geometry with a suitable
intermediate buffering region connecting the UV and IR region. Furthermore, far away from the black hole horizon,
 extremal and non-extremal
geometry become identical, which is evident from the form of the metric  presented in the last section. At large $r\gg r_h$,
$\left(\frac{r_h}{r}\right)^l\ll 1, l>0$ and thus the non-extremal metric for large $r$ indeed becomes identical to the extremal limit.  Hence
to analyze the difference between the extremal and non-extremal geometries, we will ignore the effect of UV completions in this work and
 analyze
how our results modify with the addition of various UV caps in our upcoming paper\cite{Long-Keshav} .

We will now evaluate the on-shell value of the action (\ref{Action}) for
the extremal geometry $X^1$ without adding a UV cap and any localized sources. When the axio-dilaton field
$\widetilde{\tau}$ is constant, the fluxes and the metric  were
 exactly evaluated in \cite{KS}. Thus in the absence of D7 branes such that $N_f=0,\widetilde{\tau}=$constant, we get for the {\it bulk} on shell  action
\bg \label{S1} S^1_{\rm total
}&=&\frac{1}{2\kappa_{10}^2}\left[\int
d^8x\int_0^{\beta_1}d\tau \int_{0}^\infty
d\rho\sqrt{G_1}\left(-\frac{G_3^1\cdot \bar{G}^1_3}{24{\rm
Im}\tau}\right)-i \int \frac{C_4\wedge G_3^1\wedge \bar{G}_3^1}{4 {\rm
Im}\tau
}\right]\nonumber\\
&=&0 \nd
where $G_1$ is the determinant
of the metric (\ref{metd}) after wick rotation, $G_3^1=F_3^1-\frac{i}{g_s}H_3^1$,
$H_3^1,F_3^1$ are the three form fluxes on deformed cone \cite{KS}.  Using the equations of motion for
$G_3^1$: $\ast_{10}G_3^1=iC_4\wedge G_3^1$, one readily gets (\ref{S1}) .

 Here
$\beta_1= T_1^{-1}$ can have any value corresponding to any
temperature. This is because unlike the Poincar\'e AdS geometry, the deformed cone geometry (\ref{metd}) is regular and after Wick rotation, the periodicity of Euclidean time
$\beta_1$ can take any value. For  a chosen value of $\beta_1$, $T_1$ gives
the local temperature of the field theory living on the hypersurface
$\rho=\rho_c$ for {\it any} $\rho_c$.\footnote{The four
dimensional Minkowski metric at hypersurface $\rho=\rho_c$ has
identical warp factor for both the time and space directions for
every $\rho_c$. Thus the local temperature at $\rho=\rho_c$ is
independent of $\rho_c$.} 

Now computing the on shell {\it bulk} action for the non-extremal geometry $X^2$
gives
\bg \label{S2} S^2_{\rm
total}=\frac{1}{2\kappa_{10}^2}\left[\int
d^8x\int_0^{\beta_2}d\tau \int_{\rho_h}^\infty
d\rho\sqrt{G_2}\left(-\frac{G_3\cdot \bar{G}_3}{24{\rm
Im}\tau}\right)-i \int \frac{C_4\wedge G_3\wedge \bar{G}_3}{4 {\rm
Im}\tau }\right]\nonumber\\
\nd
where now the three from flux $G_3$ is distinct
from the extremal values and is no longer ISD i.e. $\ast_6 G_3\neq iG_3 $, thus (\ref{S2}) is not zero. Also,
 $G_2$ is the determinant of the Euclidean 
metric and $\rho_h$ is the horizon in the radial coordinate $\rho$. By switching to $r$ coordinate using the transformation
(\ref{rrho}),
 the corresponding temperature of the field theory at
hypersurface $r={\cal R}$ of geometry $X^2$ is 
 \bg \label{Tncft}
T_2({\cal R})=\frac{\beta_2^{-1}}{\sqrt{g({\cal R})}}\nonumber\\
\beta_2=\frac{4\pi\sqrt{h(r_h)}}{|g'(r_h)|}
\nd
where $r_h$ is the horizon\footnote{The temperature (\ref{Tncft}) is derived by taking the non-extremal metric to be (\ref{inmated}) with
a(r)=1. That is we are assuming the horizon is large enough such that the non-extremal limit of (\ref{inmateda}) takes the form
(\ref{inmated}).} and prime denotes derivative with respect to $r$.
The two geometries describe the same field theory at the same temperature
on the hypersurface $r={\cal R}$ if

\bg \label{beta}
T=T_1({\cal R})=T_2({\cal R})\Rightarrow \beta_1=\beta_2 \sqrt{g({\cal R})}
\nd

 Also we have to take the GH term into consideration, this will be discussed in the next subsection.

Now, to see which dual geometry is favored for a gauge theory at fixed temperature $T$ and living on hypersurface ${\cal R}=\infty$,
 we consider the action difference \cite{Hawking:1982dh}\cite{Witten:1998zw}
 \bg \label{delS}
 \triangle S&=&S^2_{\rm total}+S_{\rm GH}^2-S^1_{\rm total}-S_{\rm GH}^1\nonumber\\
 &=&\frac{\beta_2}{2\kappa_{10}^2}\int dx^8\lim_{{\cal R}\rightarrow \infty}\Bigg[\int_{\rho_h}^{{\cal R}}
 d\rho \;{\cal I}_2\Bigg] \nonumber\\
  &&-\frac{i}{2\kappa_{10}^2} \int \frac{C_4\wedge G_3\wedge \bar{G}_3}{4 {\rm Im}\widetilde{\tau} }+S_{\rm GH}^2-S_{\rm GH}^1\nonumber\\
  {\cal I}_2&\equiv&\sqrt{G_2}\left(-\frac{G_3\cdot \bar{G}_3}{24{\rm Im}\widetilde{\tau}}\right)
 \nd
 where  $S^1_{\rm GH}, S^2_{\rm GH}$ are the Gibbons-Hawking surface terms for the extremal and
 non-extremal actions. If $\triangle S>0$, then extremal geometry is preferred and the dual gauge theory is in the
 confined phase. On the other hand  $\triangle S<0$ indicates that non-extremal geometry will be favored and the gauge theory is in the
 deconfined phase. Note that  $\triangle S$ only depends on $\rho_h$ and thus on  $T$. As $T$ is altered from small to large values,
 it is possible for $\triangle S(T)$ to change sign. If there exists a particular $T=T_c$ such that $\triangle S(T_c)=0$, then $T_c$ will be
 the critical temperature which is realized as the confinement/deconfinement phase transition temperature of the dual gauge theory. Since we
 are
 working in the limit $N_f=0$, the corresponding critical temperature is for pure glue large N gauge theory with no fundamental matter.

 Now to  evaluate (\ref{delS}) explicitly, we need to know $S^1_{\rm GH}, S^2_{\rm GH},{\cal I}_2$ exactly. We will now compute $S^1_{\rm GH}, S^2_{\rm GH}$ while analyzing
 ${\cal I}_2$ for large and small black holes separately and estimate the critical temperature.

 \subsubsection{Large horizons, critical temperature and deconfined phase}\label{NCFT2}
  When $\rho>\rho_h$ is
 large, the internal metric simplifies from the non-extremal limit of (\ref{metd}) and takes the form (\ref{inmated}) where the radial coordinate $r$ given by
 (\ref{rrho})
 becomes convenient. Thus for large radial distances we will use $r$ coordinate and we really
  have a Klebanov-Tseytlin type geometry. Then the three form fluxes in the
 presence of a black hole are given by  (\ref{F3}),(\ref{H3}) while the flux strength is given by (\ref{G3^2}). Furthermore, only keeping
  terms up to linear order in ${\cal O}(g_sM^2/N)$ and using $e^\phi=g_s$, ${\cal I}_2$ can be exactly evaluated with the following result

  \bg\label{I2}
  {\cal I}_2(r)&=&-\frac{3r^3 g_sM^2\left(2-\frac{\tilde{r}_h^4}{r^4}\right){\rm sin}\theta_1\;{\rm sin}\theta_2\; }{16 N}\nonumber\\
  -i \int \frac{ C_4\wedge G_3\wedge \bar{G}_3}{4 {\rm Im}\widetilde{\tau} }&=& \beta_2\int_{r_h}^{\infty} dr\; dx^8\;
  {\rm sin}\theta_1\;{\rm sin}\theta_2\; \frac{3g_sM^2r^3}{8N}
  \nd
  
To compute the GH term for KS geometry with black hole, we find the embedded Euclidean metric at the boundary,
\begin{eqnarray}\label{metric-fix-r}
ds_9^2&=&\frac{r^2}{\alpha'\sqrt{N\Big(1+\frac{81g_sM^2}{8 N}\log \frac{r}{r_l}+A_1(r)\Big)}}\Big[\Big(1-\frac{\tilde{r_h}^4}{r^4}+G(r)\Big)
d\tau^2+\vec{dx}^2\Big]\nonumber\\
&&+\,\alpha'\sqrt{N\Big(1+\frac{81g_sM^2}{8 N}\log \frac{r}{r_l}+A_1(r)\Big)}\,\,(b(r)\tilde{g}^{(1)}_{pq}dx^pdx^q)
\end{eqnarray}
where $A_1(r),G(r)$ has the form given by (\ref{gpq}).
The GH term is
\begin{eqnarray}
S_{\rm GH}=\frac{1}{\kappa_{10}^2}\int d^9x\sqrt{\textrm{det}\tilde{g}}\, g^{\mu\nu}D_{\mu}n_{\nu}=\frac{1}{\kappa_{10}^2}\int d^9x\sqrt{\textrm{det}\tilde{g}} \, g^{\mu\nu}(\partial_{\mu}n_{\nu}-\Gamma^{\alpha}_{\mu\nu}n_{\alpha})
\end{eqnarray}
where $\tilde{g}$ is the metric (\ref{metric-fix-r}). 
Since the internal metric $\tilde{g}_{pq}=k(r)\tilde{g}^{(1)}_{pq}$ is unknown, we can not determine the GH term completely. However, 
from eq. (\ref{gpq})
 we know that the corrections to $\tilde{g}_{pq}$ is of order $\mathcal{O}(g_sM^2/N)\frac{\tilde{r}_h^{4l}}{r^{4l}}, l>0$,
  so their contributions to the GH term will be of same order. Using this scaling,  we find that the GH term for KS black hole metric is
\begin{eqnarray}\label{GH2}
S^2_{GH}=\frac{1}{108\kappa_{10}^2}
\lim_{{\cal R}\rightarrow \infty}\Big[4{\cal R}^4-2\tilde{r}_h^4+\frac{729g_sM^2}{16 N}({\cal R}^4-(1+d)\tilde{r}_h^4)\Big]\beta_2V_8\nonumber\\
\end{eqnarray}
where $V_8\equiv V_3\times V_5$. Also the term $\sim d~\tilde{r}_h^4g_sM^2/N$ 
comes from the 
$\tilde{g}'_{pq}, A_1(r)$ and $G(r)$ corrections that modifies the unit normal vector $n^\mu$ and its covariant derivative. In deriving (\ref{GH2}), we only kept terms up to linear order in $\frac{g_sM^2}{N}$ which is
valid when $N\gg M, N\gg {\rm log}\left(\frac{{\cal R}}{r_l}\right)$. When there is no black hole, i.e $r_h=\tilde{r}_h=0$, we get the GH term for KS geometry.
That is 
\begin{eqnarray}\label{GH}
S^1_{GH}=\frac{1}{108\kappa_{10}^2}\lim_{{\cal R}\rightarrow \infty}{\cal R}^4\Big[4+\frac{729g_sM^2}{16 N}\Big]\beta_1V_8\nonumber\\
\end{eqnarray}

Then we get the action difference
  \bg \label{delSa}
  \triangle S&=&\frac{3g_sM^2\beta_2 V_8r_h^4}{32\kappa_{10}^2N}\lim_{{\cal R}\rightarrow \infty}
  \left({\rm log}\left(\frac{{\cal R}}{\tilde{r}_h}\right)-\frac{9}{4}-\frac{9}{2} \left[d-\alpha^1\right]\right)
   \nd
where $\alpha^1$ arises from the following expansion  
\bg
e^{2B}=g(r)\equiv 1-\frac{\tilde{r}_h^4}{r^4}\left(1+
 \frac{729 \alpha^1 g_sM^2}{8 N}\right)+ \frac{\tilde{r}_h^8}{r^8} {\cal O}\left(\frac{g_sM^2}{N}\right)+...
\nd

Due to the presence of D5 branes, we expect the black hole horizon to be larger than the Schwarzschild value \cite{fangmia}, which is possible if
$\alpha^1>0$. In the rest of the analysis, we will assume $\alpha^1>0$. 
Note that $d$ is determined by the Einstein
equations and the flux equations once boundary conditions are imposed. Since $d$ arises from the coefficients of the term 
$\frac{\tilde{r}_h^4}{r^4}$ in the expansion (\ref{gpq}) of 
$\tilde{g}'_{pq}, A_1(r), G(r)$, it is sensitive to the near horizon geometry and in particular can be obtained from the horizon values of
these functions. Furthermore, $d$ is related to $A_1(r_h)$, which in turn determines the number of effective degrees of freedom at a
temperature $T\simeq r_h/L^2$. Thus from the gauge theory side, $d$ is related to the effective colors at the thermal scale.  
As we are not able to exactly solve all the Einstein equations near
the black hole horizon, we can speculate two scenarios: 

\vspace{15pt}

\noindent $\bullet$ $d\le \alpha^1-1/2$: In this case $\triangle S> 0$, which means extremal geometry is favored and black hole is not formed. In that
 case all black holes
have higher free energy and for all corresponding temperatures, the gauge theory will be in the confined  phase. 
However,  this does not mean there is no phase transitions at all- rather upto linear order in $\frac{g_sM^2}{N}$, $\triangle S> 0$. 
In fact there are
terms in $\triangle S$ which are second or higher order in $\frac{g_sM^2}{N} {\rm log}\left(\frac{{\cal R}}{r_l}\right)$ and for 
${\cal R}\rightarrow \infty$, 
it is possible
that $\frac{g_sM^2}{N} {\rm log}\left(\frac{{\cal R}}{r_l}\right)\gg 1$.  Then the perturbative analysis breaks down and we cannot ignore higher order terms. When the higher order terms are accounted for, 
$\triangle S$ could indeed be zero for some $r_h^c$ and there would be a phase transition. 

\vspace{15pt}

 \noindent $\bullet$  $d> \alpha^1-1/2$: In this case, using (\ref{delSa}), it is possible to obtain $\triangle S=0$ with the following value for 
 critical horizon
\bg\label{Tc}
\tilde{r}_h^c&=&\frac{\mathcal{R}}{{\rm  exp}\left(\frac{9}{4}\left[1+2(d-\alpha^1)\right]\right)}
\nd
There is always a phase transition provided $\frac{g_sM^2}{N} {\rm log}\left(\frac{{\cal R}}{r_l}\right)\ll 1$. This is possible with the
 scaling ${\cal R}=N^{1/4}\sqrt{\alpha'}\rightarrow \infty$, $r_l\sim {\cal O}(\sqrt{\alpha'})$ and  $N\gg M^2$. Then the 
 corresponding critical temperature is
\bg\label{Tcc}
T_c&=&\frac{1+{\cal O}\left(\frac{g_sM^2}{N}\right)}{\pi \;{\rm 
exp}\left(\frac{9}{4}\left[1+2(d-\alpha^1)\right]\right)N^{1/4}\sqrt{\alpha'}}\nonumber\\
&\sim& \frac{g_s^{1/4}}{\lambda^{1/4} \sqrt{\alpha'}}
\nd
 where $\lambda\equiv g_sN$ is the 'tHooft coupling. For $T>T_c$, the black hole is formed and describes the deconfined
phase which has non-zero entropy. On the other hand for $T<T_c$, we have a confined phase with zero entropy described by the extremal geometry.
 Thus at $T=T_c$, there is a first order phase
transition in the gauge theory and we have obtained a gravitational description of it in terms of Hawking-Page phase transition between two
geometries.  

Now to obtain thermodynamic state functions of the gauge theory, we first need to renormalize the action.
From eq. (\ref{GH2}), we see that the on shell action is divergent and we need the following counter terms: 
\bg
S^2_{\rm counter}=-\lim_{{\cal R}\rightarrow \infty}{\cal R}^4 \frac{\beta_2 V_8}{\kappa_{10}^2}\left(
\frac{1}{27}+\frac{27g_sM^2}{64N}+ \frac{\kappa \tilde{r}_h^4}{ 54{\cal R}^4}\right)
\nd
where $\kappa$ can be determined by matching the Bekenstein-Hawking entropy of the black hole to the entropy obtained through the partition
function using the relations (\ref{KS16}) and  (\ref{pE}).
 With this counter term, for $T>T_c$, we can write down the free energy and entropy 
for the deconfined phase of non-conformal field theory using (\ref{FreeE}, \ref{pE})
\bg\label{FreeEn}
F&=&-\frac{V_8 \tilde{r}_h^4}{2\kappa_{10}^2}
\left[\frac{1+\kappa}{27} -\frac{3g_sM^2}{16N}
  \left({\rm log}\left(\frac{{\cal R}}{\tilde{r}_h}\right)-\frac{9(1+d)}{2}\right)\right]\nonumber\\
  &\sim&
  -N^2T^4\left(1+ \frac{g_sM^2b }{N} {\rm log}(LT)\right)\nonumber\\
  s&\sim& N^2 T^3\left(1+ \frac{g_sM^2b }{N}\left(1+ {\rm log}(LT)\right)\right)
\nd
where $b$ is some constant independent of $M$, $N$ but determined by the relation between $\tilde{r}_h$ and $T$. We observe that  
$-F/\tilde{r}_h^4>0$ for $N\gg M$ and increases with $\tilde{r}_h$. This indicates entropy increases with $\tilde{r}_h$. Also temperature $T$ is
given by (\ref{Tncft}) and since the exact metric at the horizon is not known, the exact expression for $T$ as a function of
$\tilde{r}_h, M$ and  $N$ is also unknown. This is the reason the exact scaling of free energy with temperature is unknown and in 
(\ref{FreeEn}), we
approximated $T\sim \frac{\tilde{r}_h}{\pi \sqrt{N}\alpha'}$, which is a reasonable since it is the leading term and $g_sM^2/N \ll 1$. 
Using the same renormalization procedure for the extremal action gives 
$S^{1}_{\rm counter}= S^2_{\rm
counter}(\beta_1\rightarrow \beta_2, \tilde{r}_h=0)$ and we get $F=0,s=0$ for $T<T_c$ - the confined phase. 

The above analysis, done purely using supergravity can be interpreted 
as a possible proof of deconfinement at large temperatures for a gauge theory which has a gravity dual.
On the other hand, when temperature is zero we have
the unique extremal geometry corresponding to the confined phase. Thus starting with a  confining phase at zero temperature, our
analysis shows that the gauge theory  transitions to a  deconfined phase provided  $d> \alpha^1-1/2$ i.e the deformations of $T^{1,1}$ near the horizon
behave in a particular way. 
Thus, we have a
confinement/deconfinement phase transition and the entire analysis is done using the low energy limit of type IIB superstring action
(\ref{Action}).

\subsubsection{Small horizons and the confined phase}
From (\ref{delSa}),  observe that  as long as $\tilde{r}_h<\tilde{r}_h^c $, $\triangle S> 0$ and the extremal geometry without a black hole
will be preferred over the black holes geometry. Thus when the horizon radius is small compared to the critical radius
$\tilde{r}_h^c$, the dual gauge theory will be in the confined phase.

Note that our entire analysis has been done in the large $\rho$ region where the relevant radial coordinate is $r$. Going back to
the $\rho$ coordinates, we see that the critical horizon in terms of $\rho$ coordinate becomes
\bg \label{rhoc}
\rho_c\sim 3{\rm log}\left(\frac{\tilde{r}_h^c}{{\cal A}^2}\right)
\nd
 The horizon
$\rho_h\sim 3{\rm log}\left(\frac{\tilde{r}_h}{{\cal A}^2}\right) <\rho_c$ can still be a large number referring to a large black hole. But even
this `large' black hole is not favored against the extremal solution since $\rho_h<\rho_c$.
Thus critical horizon is the relevant scale that determines largeness and smallness.

For $\rho_h\ll \rho_c$, we do not have explicit form of the metric and the flux and thus cannot evaluate the on shell non-extremal action $S^2$.
Equation (\ref{delSa}) does not apply in such cases and we need explicit solutions of the metric and fluxes to
evaluate the action difference. Note that for these small black holes, we have to consider the non-extremal limit of (\ref{metd}), which is not known.
The difficulty arises for the following reason: the proposed solution (\ref{inmated}) with the scalings (\ref{gpq}) is only valid for
large black holes. In finding these scalings, we heavily
relied on the smallness of the parameter $g_sM^2/N$. For small black holes, such scalings do not apply since at small $\rho$, $N$
cascades down to smaller and smaller values. The relevant quantity now is $g_sM^2/N_{\rm eff}(\rho)$ which is no longer small and the
perturbative analysis can no longer be applied.

However,  using (\ref{delSa}) which is only applicable for large $\tilde{r}_h$ i.e large $\rho_h$, we observe that
 $\triangle S=0$ only once. On the other hand small $\rho_h$ means small temperature and in particular  at zero
 temperature, the $SU(M)$ pure glue dual gauge theory confines.
 Since there is nothing special that occurs in the gauge theory at low temperature, it is reasonable that $\triangle S\neq 0$
 at low temperature. This implies $\triangle S$ becomes zero only at high temperatures and
 we have unique critical temperature  given by the critical horizon $\rho_c$ in
 (\ref{rhoc}). For
 every $\rho_h<\rho_c$, no matter how small it is, we will have confinement. Of course this is consistent with confinement at zero temperature
 and the scenario that deconfinement happens only at a single large temperature- indicating we have only two phases, confined and the
 deconfined phase. To gain a better understanding  of the gauge theory and its precise degrees of freedom in the different phases,
 we will now sketch a possible brane configuration that gives rise to the dual geometries $X^1$ and $X^2$.

\section{Brane configurations, the gauge theory and connection to QCD}\label{Brane}
In the previous sections we studied two gravity solutions and the
Hawking-Page like phase transition between them, in this section we will
study the brane configurations that lead to the dual gauge theories.

Suppose we place N D3 branes  at the tip of a regular cone [See Fig
1(a)]. At zero temperature, the gauge group is $SU(N)\times SU(N)$
with bi-fundamental fields $A_i,B_j, i,j=1,2$. This gauge theory has
a conformal fixed plane and the number of D3 branes remains the same
at all energy scales. Now if we put another stack of D5 branes that
wraps the vanishing two cycle at the tip of the cone [See Fig
1(b)] the gauge
theory becomes $SU(N+M)\times SU(N)$ with the bi-fundamental fields,
and it is no longer conformal. The $SU(M+N)$ sector has $2N$
effective flavors while the $SU(N)$ sector has $2(N+M)$ effective
flavors thus it is dual to the $SU(N-M)\times SU(N)$ gauge theory
under an Seiberg duality. Under a series of such dualities which is
called cascading, at the far IR region the gauge theory can be
described by $SU(M)\times SU(K)$ group, where $N=lM+K$, $l$, $0\leq
K<M$ are positive integers. Now the number of `actual' D3 branes N
is no longer the relevant quantity, rather $N\pm pM$
where $p$ is an integer describes the D3 brane charge. We take $K=0$ in all our analysis, so at
the bottom of the cascade,
 we are left with  ${\cal N}=1$ SUSY $SU(M)$ strongly coupled
 gauge theory which looks very much like strongly coupled SUSY QCD.
\begin{figure}[htb]\label{F1}
                \begin{center}
		\vspace{- 0.5 cm}
		\centering
		\subfloat[N D3 branes placed at the tip of the conifold]{\includegraphics[width=0.33\textwidth]{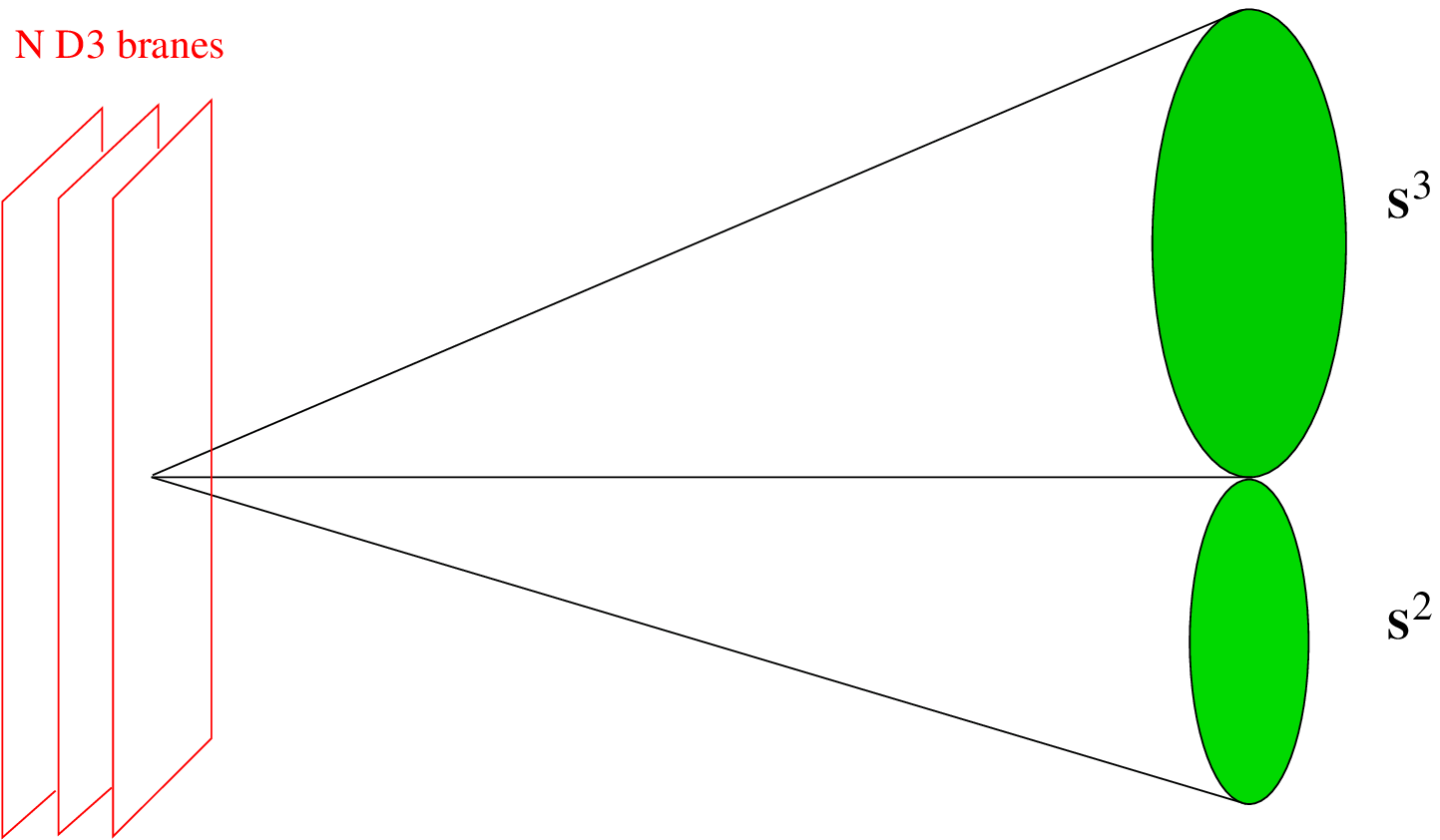}}
		\subfloat[M D5 branes wrapping the vanishing two cycle of the conifold]{\includegraphics[width=0.33\textwidth]{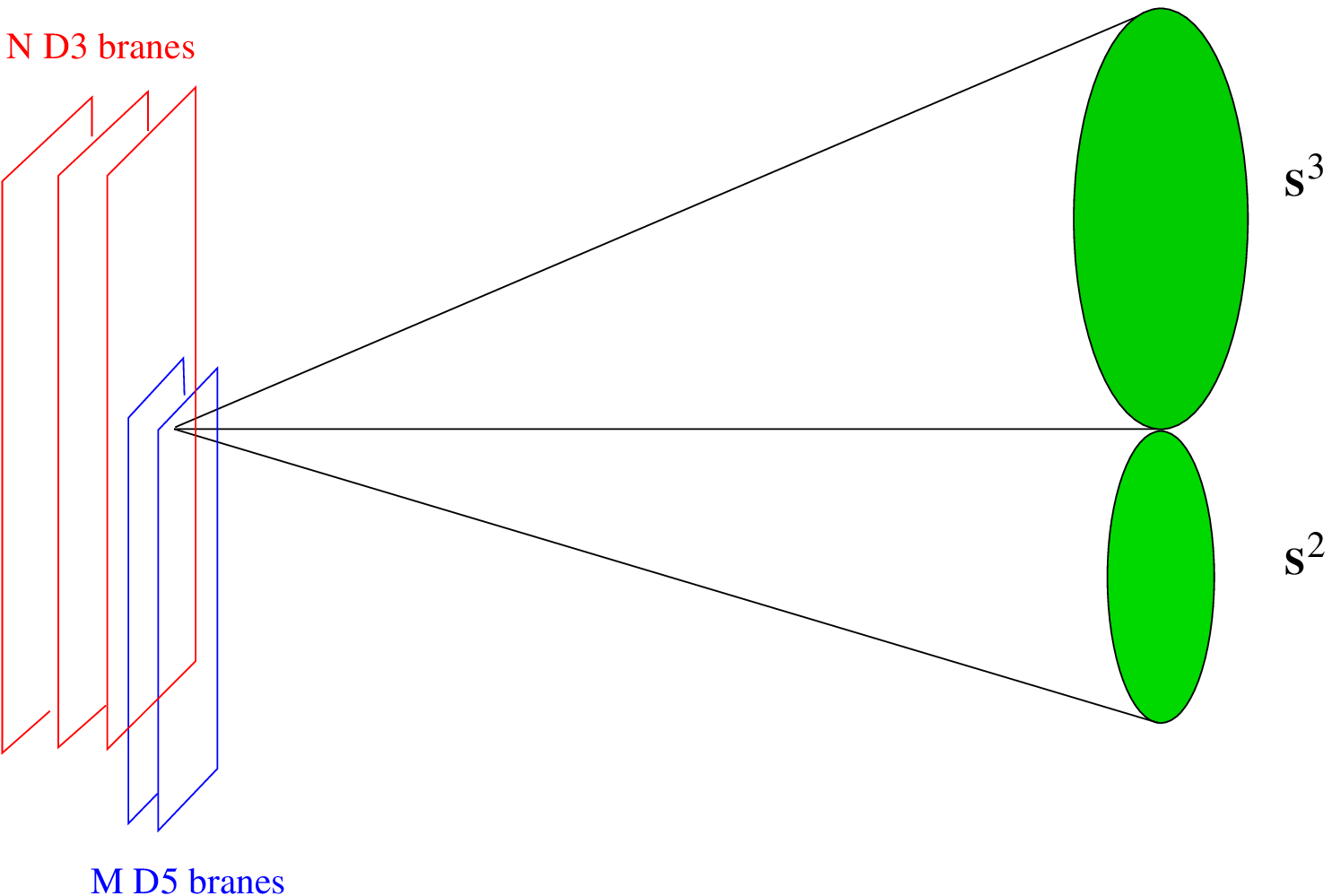}}
        \subfloat[M Anti-5 branes separated from the D5 and D3 branes but also located at the tip $r=0$.]{\includegraphics[width=0.33\textwidth]{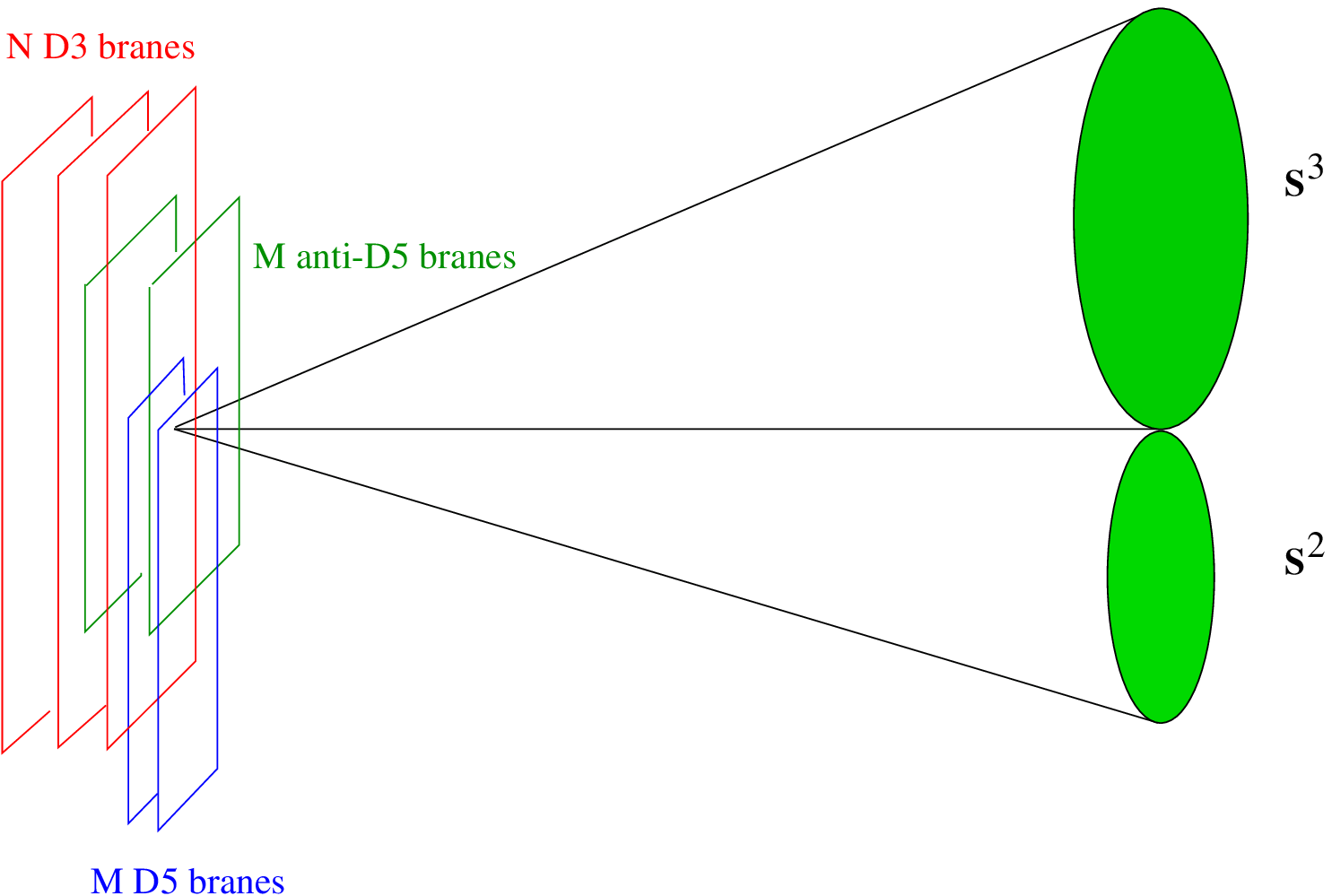}}
		\caption{{Brane construction of conformal field theory, non-conformal field theory and UV completed non-conformal field theory }}
\end{center}
\end{figure}

 Due to the strong coupling at the IR, the gauge theory develops non-perturbative superpotential \cite{Affleck:1983mk} and  breaks the
 $Z_{2M}$ R symmetry  down to $Z_2$ group. Since the complex fields $A_i,B_j, i,j=1,2$ also describe the complex coordinates of the cone,
 the breaking of the R symmetry  modifies the geometry from a regular to a deformed cone \cite{KS}. Thus to capture the IR modification
 of the gauge theory, we must consider the {\it warped deformed cone}.

All the discussions so far have been concerned with zero temperature
dynamics of the gauge theory. At finite temperature, we expect the
number of D5 branes $M$ remains constant at different temperatures,
thus on the gravity side the units of $F_3$ fluxes remains the same
too. The three form flux $F_3$ is given by (\ref{F3}). {\it This is
the crucial step in our analysis since in the presence of the black
hole, the base $T^{1,1}$ is modified and the spheres $S^2, S^3$ are
now squashed
 and deformed. But we assume that at $r\mapsto \infty$ the base is
 still $T^{1,1}$ and the $S^3$ at the infinity still encloses the
 directions which the D5 branes extend along on the gauge theory side.
 Thus the integral of the RR flux $F_3$ over this $S^3$ should be
 equal to the the number of D5 branes $M$ and $F_3$ is the same as
 in the extremal case.}

Although $F_3$ remains the same with or without the black hole\footnote{$F_3$ remains the same but $\tilde{F}_3=F_3-C_0 H_3$ changes in the
presence of the black hole, which is consistent with our earlier analysis \cite{fangmia}.}, $H_3$
given by (\ref{H3}) up to linear order in ${\cal O}(g_sM^2/N)$ is
distinct from the extremal flux, $H_3^0= \frac{3g_sM\alpha'}{2r}
dr\wedge \omega_2$. So the three form flux $G_3$ in the presence of
the black hole is modified and it is no longer ISD. Since $H_3$ is
modified, $B_2$ is also modified and this means that the change in
$B_2$ can alter the RG flow. But as discussed in \cite{fangmia},
this modification does not alter how the gauge theory couplings flow
with energy scale, it only redefines the energy scale in terms of
temperature.

Now observe that the gauge group becomes $SU(M)$ only at the far IR
while at high energies, it can be described by
$SU(k(\Lambda)M)\times SU((k(\Lambda)-1)M)$ group, with $k(\Lambda)$
increasing with energy. Thus the UV of the gauge theory has
divergent effective degrees of freedom and looks nothing like QCD.
Although the confined phase of the gauge theory may resembles
${\cal N}=1$ SUSY QCD, the
deconfined phase of the gauge theory is quite different. Hence the
value for critical temperature (\ref{Tcc}) cannot be directly related
to the confinement/deconfinement temperature of QCD.

As already discussed in section \ref{extm}, to  make connections
to large $N$ QCD, we must add localized sources that can modify the UV
dynamics of the gauge theory and consequently alter the large $r$
region of the dual geometry. One possibility is to add $M$ anti-five
branes separated from each other and  from the $D3-D5$ branes at the tip of the cone. To obtain this separation, we must blow up one of the
$S^2$'s at the tip and give it a finite size - which essentially means putting a resolution parameter. 
 The setup is sketched in Fig. 1(c) [we did not blow up the tip, which was intentional and will be explained in what follows]. 
 The separation gives masses $\Lambda_0$ to the $D5-\bar{D}5$ strings and at scales less than the mass, the
 gauge group is $SU(N+M)\times SU(N)\times U(1)^M$ where the additional $U(1)$ groups arise due to the massless strings ending on the same 
 $\bar{D}5$
 brane.  At scales much larger than $\Lambda_0$, $D5-\bar{D}5$ strings are excited and we have $SU(N+M)\times SU(N+M)$ gauge theory. 
 For $\Lambda<\Lambda_0$, i.e. at  low energy, gauge theory is best understood as arising  from the set up of Fig. 1(b) (since the
 modes from 
$\bar{D}5$ branes are not excited). At high energy $\Lambda\gg \Lambda_0$, the gauge theory is best described as arising from  
Fig. 1(c).
Essentially at high energies, the resolution $r_0\sim \Lambda_0$ is not significant and  we have effectively 
$N+M$ D3 charge at the tip of the
regular cone, which arises by placing $M$ number of $D5-\bar{D}5$ pairs at the tip.

Of course there will be tachyonic modes arising from $D5-\bar{D}5$ strings and one needs to consider electric and magnetic fluxes on the
respective branes to stabilize the system \cite{susyrest1}-\cite{susyrest7}. Furthermore, to incorporate fundamental matter and chiral symmetry
breaking, we need to add $D7-\bar{D}7$ branes to the brane configuration of Fig. 1 and the details were discussed in \cite{Chen:2012me}. 
The RG flow arising from a UV complete model with $D3,D5$ and anti $D5$ branes in sketched in  the Fig.2.
\begin{figure}[htb]\label{gauge}
       \begin{center}
\includegraphics[height=9cm]{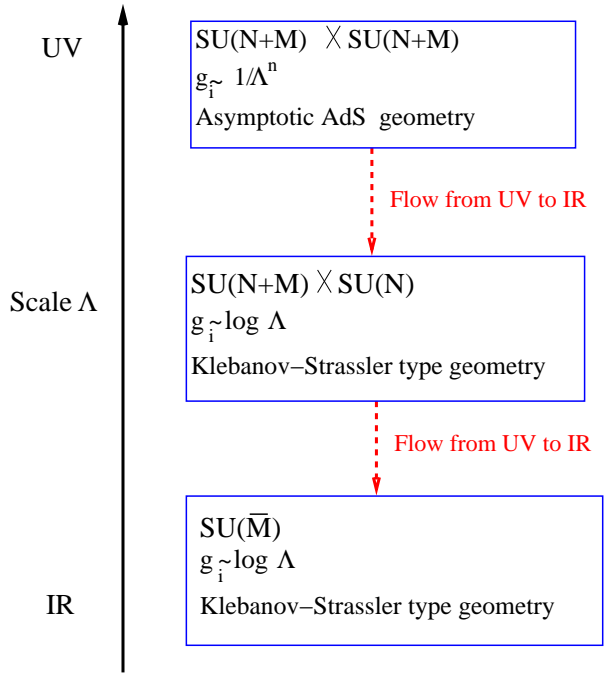}
        \caption{Gauge theory arising from the brane configuration in Fig. 1(c)}
       \end{center}
        \end{figure}

On the supergravity side, the anti five branes will source RR and NS-NS  three form flux such
that the total flux $F_3,H_3\rightarrow 0$ as $r\rightarrow \infty$.
As already discussed in section \ref{extm}, vanishing of the
fluxes indicate that the gauge theory reaches conformal fixed point
where the two gauge couplings $g_1,g_2$ becomes identical. At the
far UV, since both the groups have same rank, we effectively have a
single Yang Mills coupling $g_{\rm YM}\rightarrow 0$ with 'tHooft
coupling $g_{\rm YM} N_{\rm eff}$ held fixed at large value. Thus
this UV completion arising from anti brane sources in principle
gives rise to a QCD like theory which confines in the IR and becomes
conformal in the UV. In fact the Yang Mills coupling of the gauge
theory becomes free in the far UV- just like QCD, indicating that
the gauge theory is indeed very similar to large $N$ QCD.

\section{Conclusions}

In this paper we studied the non-extremal geometries on the dual gravity side of the non-conformal finite temperature field theory.  
To find the exact non-extremal geometry analytically is extremely difficult because the solution is no longer supersymmetric
 and we have no control over 
the internal manifold. Even numerically it will be a formidable task to solve all the Einstein equations and field equations together. 
However up to linear order in $O(g_sM^2/N)$, the supergravity equations drastically simplify and the deformations of the internal metric along
with the corrections to black hole factor and the warp factor can be evaluated as an infinite Taylor series in $\frac{r_h}{r}$.
Using this expansion, it is straight forward to evaluate the on shell ten dimensional gravity action and
 we find a  phase transition between the 
extremal  and non-extremal geometry. Since the geometry is dual to a non-conformal gauge theory, this transition is  interpreted
  as the confinement/deconfinement phase transition.

The non-extremal gravity solution we discussed is not UV completed. However, under the scaling ${\cal R}=N^{1/4}
\sqrt{\alpha'}\rightarrow \infty$ and
$N\gg M$, we get $\frac{g_sM^2}{N} {\rm log}\frac{{\cal R}}{r_l}$ to be small and there is no logarithmic divergence. In fact there are no
${\rm log}({\cal R})$ divergence in the Gibbons-Hawking boundary term. The only term that has logarithmic term arises from 
 the bulk action- but since $\tilde{r}_h>\tilde{r}_h^c$ scales with ${\cal R}$, the term ${\rm log}\frac{{\cal
R}}{\tilde{r}_h}$  is in fact convergent. 
 The critical temperature of the confinement/deconfinement we find in section \ref{PT} is obtained using   this particular scaling and thus
 our analysis is somewhat restrictive.  
 It will be interesting to UV complete the geometry and then there will be no need for this particular scaling since there will no logarithmic
 divergences to begin with. Using the UV complete geometry to study the phase transition and other 
 thermal properties of the 
 non-conformal field theory is the content of our upcoming work \cite{Long-Keshav}. 
 
 We also did not include effect of running dilaton in the black hole geometry and did not consider D7 branes and other localized sources
 in non-extremal dual geometry. This means the thermodynamics we obtain is strictly restrictive to gauge theory with no flavor, no
 Baryochemical potential  and it is not
 surprising that we obtain a first order phase transition. However the perturbative procedure outlined here to solve the Einstein equations
 along with the flux equations can easily be generalized in the presence of a running dilaton field and other localized sources. Similarly the
 on shell action in the presence of running axio-dilaton fields and localized sources can also be evaluated up to linear order in our
 perturbative parameter. Note that for a UV complete scenario, we must replace $M\rightarrow M(r),
  \lim_{r\rightarrow \infty} M(r)\rightarrow 0$ and thus the on shell value of the action will be significantly different. In fact, the
  presence of localized sources will dictate how fast $M(r)$ vanishes- which also indicates how rapidly the theory becomes conformal. This
  means the width of the conformal anomaly will be highly sensitive to the details of the localized sources. In \cite{Chen:2012me} a detailed
  gauge theory analysis was done and  the non-extremal dual geometry will be presented in \cite{Long-Keshav}      
\vskip.2in

\centerline{\bf Acknowledgement}

\noindent We would like to especially thank Keshav Dasgupta for  explaining the brane configuration and various discussions
during the course of the work and Miklos Gyulassy for his
valuable feedback. We would also like to thank Christopher Herzog for helpful discussions. The work of M. M. is supported in part by the Office of Nuclear Science of the US
Department
of Energy under grant No. DE-FG02-93ER40764.
\appendix
\section{Appendix} \label{apdx}
In this appendix we study $\alpha=e^{4A+B}$ case in section $2$, assuming the internal 5D manifold $\mathcal{M}$ is $T^{1,1}$ and $G_3=0$.

By simplifying the Einstein equations and flux equations we finally get the following four equations,
\begin{eqnarray}
k^5e^{-4A+B}(4A'+B')&=&1\nonumber\\
2B''+4B'^2+5\frac{k''}{k}-\frac{a'}{a}B'-\frac{5}{2}\frac{a'}{a}\frac{k'}{k}+10\frac{k'}{k}B'&=&0\nonumber\\
2e^{2B}akk''+8e^{2B}ak'^2+4e^{2B}akk'B'+4e^{2B}ak^2B'^2&&\nonumber\\
-e^{2B}a'k^2B'-e^{2B}a'kk'-8a^2+2e^{2B}ak^2B''&=&0\nonumber\\
\frac{a'}{a}-10\frac{k'}{k}-4B'-2\frac{B''}{B'}&=&0
\end{eqnarray}
After further simplification we find,
\begin{eqnarray}
&&k=\frac{1}{(1-e^{2B})^{1/4}}, \quad\quad a=e^{2\int \frac{dk'}{k}}, \nonumber\\
&&B'+2-2e^{-2B}=0, \quad\textrm{or}\quad B'^2(1+4e^{-2B})=2B''(1-e^{-2B})
\end{eqnarray}
The first differential equation of $B$ gives $e^{2B}=1-1/r^4$ which is the usual AdS black hole solution. The second differential equation of $B$ can be simplified as
\begin{eqnarray}\label{Bd}
B'=e^B(-1+e^{-2B})^5/2
\end{eqnarray}
It does not have an analytic expression, but we can see that $e^{2B}\leq 1$. When $B\ll 0$ eq. (\ref{Bd}) can be approximately solved $e^{2B}\simeq 4r$. Once $B$ is known we can easily get $A$, $a$ and $k$.

Now if $G_3$ is non-vanishing it should be ISD as in the GKP paper, and thus it does not effect the internal manifold at all, the only thing that changes is the Bianchi identity of the five form flux, and thus the warp factor $A$. So we have another solution with the same $a$, $k$ and $B$ but different $A$.

We are not very sure about the use of this kind of solutions or whether they are gravity duals to some field theory. It might be interesting to get numerical solutions and explore the properties of them such as stability, KK reductions, etc.

\end{document}